\documentclass[twocolumn]{aastex63}

\usepackage{comment}
\usepackage{ulem}
\usepackage{tikz}
\usepackage{color}
\usepackage{graphicx}
\usepackage{amsmath}
\usepackage{amsfonts}
\usepackage{outlines}
\usepackage{enumitem}
\usepackage{relsize}

\newcommand{\beq}{\begin{equation}}
\newcommand{\eeq}{\end{equation}}
\newcommand{\bea}{\begin{eqnarray}}
\newcommand{\eea}{\end{eqnarray}}

\newcommand{\gv}[1]{\ensuremath{\mbox{\boldmath$ #1 $}}} 
\newcommand{\trm}[1]{\textrm{#1}}

\newcommand{\non}{\nonumber \\}

\newcommand{\numax}{\nu_{\rm max}}
\newcommand{\muHz}{\trm{ $\mu$Hz}}

\DeclareGraphicsExtensions{.png,.pdf}

\begin{document}

\shortauthors{}
\shorttitle{}

\title{\bf \large Damping of Oscillations in Red Giants by Resonant Mode Coupling}

\correspondingauthor{Nevin N. Weinberg}
\email{nevin@uta.edu}

\author[0000-0001-9194-2084]{Nevin N. Weinberg}
\affiliation{Department of Physics, University of Texas at Arlington, Arlington, TX 76019, USA}

\author[0000-0001-5611-1349]{Phil Arras}
\affiliation{Department of Astronomy, University of Virginia, P.O. Box 400325, Charlottesville, VA 22904, USA}

\author[0000-0003-4774-1351]{Debaditya Pramanik}
\affiliation{Department of Physics, Massachusetts Institute of Technology, Cambridge, MA 02139, USA}

\defcitealias{Weinberg:19}{WA19}


\begin{abstract} 
Asteroseismic studies of red giants generally assume that the oscillation modes can be treated as 
linear perturbations to the background star.  However, observations by the {\it Kepler} mission show that the 
oscillation amplitudes increase dramatically as stars ascend the red giant branch. The importance of nonlinear 
effects should therefore be assessed.  In previous work, we found that mixed modes in red giants are unstable to 
nonlinear three-wave interactions over a broad range of stellar mass and evolutionary state.  Here we solve the 
amplitude equations that describe the mode dynamics for large networks of nonlinearly coupled modes.  The networks 
consist of stochastically driven parent modes coupled to resonant secondary modes (daughters, granddaughters, 
etc.).  We find that nonlinear interactions can lower the energy of gravity-dominated mixed modes by $\gtrsim 80\%$ 
compared to linear theory. However, they have only a mild influence on the energy of pressure-dominated mixed 
modes.  Expressed in terms of the dipole mode visibility $V^2$, i.e., the summed amplitudes of dipole modes relative to 
radial modes, we find that  $V^2$ can be suppressed by $50-80\%$ relative to the linear value for highly-evolved red giants whose frequency of maximum power $\nu_{\rm max} \lesssim 100\muHz$.  However, for less evolved red giants with 
$150\lesssim \nu_{\rm max} \lesssim 200\muHz$, $V^2$ is suppressed by only $10-20\%$.  We conclude that resonant 
mode coupling can have a potentially detectable effect on oscillations at $\numax \lesssim 100\muHz$ but it cannot 
account for the population of red giants that exhibit dipole modes with unusually small amplitudes at high 
$\numax$.
\vspace{1.1cm}
\end{abstract}

\section{\bf I\lowercase{ntroduction}}
\label{s:intro}

Turbulent motions in the convective envelopes of stars excite a rich spectrum of oscillation modes.  The Sun is the best-studied example, as surface motions caused by acoustic waves can be observed with high spatial resolution (see, e.g., the review by \citealt{CD:02}). More recently, ground and space-based telescopes, particularly the \emph{CoRoT}  \citep{Baglin:06},  \emph{Kepler} \citep{Borucki:10}, and \emph{TESS} missions \citep{Ricker:15}, have observed convectively-excited acoustic and mixed modes in stars on the sub-giant and red-giant-branch (RGB).  These  observations have yielded a wealth of information about the internal and global properties of thousands of stars (see reviews by \citealt{Hekker:17, Aerts:19, Basu:20}).

In most studies, the oscillation modes are treated as small perturbations to the stellar structure, as described by the linearized fluid equations (\citealt{Aerts:10}). 
In this approximation, the modes of solar-like oscillators can only be driven by turbulent convection and damped by linear processes (e.g., turbulent viscosity and radiative damping). Beginning at lowest nonlinear order, however, the modes can interact with each other,  providing a new source of driving and damping.

In the case of the Sun,  high angular degree $p$-modes  experience three-wave nonlinear coupling with a large number of triplets composed of $f$- and $p$-modes \citep{Kumar:89}. The coupling occurs predominantly near the top of the convection zone and the overlying isothermal layer, where the wave amplitudes are largest.  The nonlinear energy transfer among trapped waves attempts to establish an equipartition of mode energies.  Furthermore, coupling to traveling waves above the acoustic cutoff frequency leads to damping of the trapped waves. In addition to modifying the mode amplitudes, these processes may broaden the linewidths \citep{Kumar:89} and induce frequency shifts \citep{Kumar:94}.   These  studies nonetheless find that nonlinear interactions in the Sun have only a modest effect on the mode amplitudes, linewidths, and frequencies.

Nonlinear interactions may be more important in RGB stars since the mode amplitudes are significantly larger than those in the Sun. By characterizing the power excess of $\simeq 1200$ \textit{Kepler} red giants,  \citet{Mosser:12:depressed}  showed that the bolometric oscillation amplitudes on the RGB are $\sim 10-100$ times larger than the Sun's, and increase dramatically as the stars evolve up the RGB (see also \citealt{Vrard:18}).  The amplitudes are larger because the convective motions are especially vigorous in the low density envelope of red giants (see, e.g., \citealt{Kjeldsen:95, Samadi:07}). 

In addition to larger mode amplitudes, the large Brunt-V{\"a}is{\"a}l{\"a} frequency in the cores of RGB stars allows them to support mixed modes, which have an acoustic character in the convective envelope and a gravity wave character in the radiative core. In our previous study (\citealt{Weinberg:19}; hereafter WA19), we showed that mixed modes in RGB stars may experience significant nonlinear coupling at the center of the star (rather than at the surface, where it occurs for solar $p$-modes).  The steepening of the waves as they approach the center  can lead to wave breaking for sufficiently low frequency waves, or to weakly nonlinear three-wave energy transfer for somewhat higher frequency waves. Previously, weakly nonlinear wave interactions in the radiative core of Sun-like stars have been studied primarily in the context of tidal friction (e.g., \citealt{Kumar:96, Barker:11, Weinberg:12, Essick:16}; but see also \citealt{Press:81} for an analysis of $g$-modes in the Sun).

The interaction of mixed modes at the center of an RGB star further differs from the interaction of  $p$- and $f$-modes in the Sun in that the mixed modes are susceptible to the {\it stochastic parametric instability}.  This occurs when a turbulently driven ``parent" mode destabilizes two resonant ``daughter" modes of infinitesimal amplitude, causing them to grow exponentially (in the Sun, the three-wave interactions instead take the form of non-resonant inhomogeneous forcing; \citealt{Kumar:89}). If the daughters' amplitudes approach that of the parent, they can act to damp the parent and thereby decrease its energy and broaden its linewidth. Since the daughters may themselves excite additional generations of modes, the saturation of the instability may involve the transfer of energy among a large network of nonlinearly interacting oscillation modes.

A principal motivation for our study is the observation of red giants with dipole mode amplitudes that are suppressed relative to the neighboring radial mode amplitudes  \citep{Mosser:12:depressed, Stello:16a, Mosser:17}. \citet{Fuller:15}  have argued that the suppression is due to the ``magnetic greenhouse effect", in which dipole mixed modes scatter off a magnetized core and get trapped deep within the star (see also \citealt{Loi:18}). However, \citet{Mosser:17} found that the observations may not be entirely consistent with the predictions of the magnetic greenhouse model. We are interested in evaluating whether nonlinear energy transfer in the core can provide an alternative explanation. 

In this study, we carry out a comprehensive analysis of the saturation of the stochastic parametric instability over a broad range of stellar mass and evolutionary state. In Section~\ref{sec:method}, we present our calculational method and discuss the key parameters that govern nonlinear mode interactions in RGB stars. In Section~\ref{sec:networks}, we describe our procedure for constructing the networks of interacting modes and in Section~\ref{sec:results} we present the results of integrating these networks. In Section~\ref{sec:observations}, we compute the visibility of the parent dipole modes as a function of stellar mass and evolutionary stage  and compare these results to the dipole mode visibilities measured with \emph{Kepler}.  We summarize and conclude in Section~\ref{sec:summary}.

\section{\bf C\lowercase{alculational} M\lowercase{ethod}}
\label{sec:method}

The Lagrangian displacement  $\gv{\xi}(\gv{x}, t)$, which relates the position $\gv{x}$ of a fluid element in the unperturbed star  to its position in the perturbed star $\gv{x}' = \gv{x} + \gv{\xi}(\gv{x}, t)$ at time $t$, satisfies the second-order equation of motion  
\beq
\label{eq:xieqn}
\rho \ddot{\gv{\xi}}= \gv{f}_1[\gv{\xi}]+ \gv{f}_2[\gv{\xi},\gv{\xi}],
\eeq
where overdots denote time derivatives, $\gv{f}_1[\gv{\xi}]$ are the linear forces, and $\gv{f}_2[\gv{\xi}, \gv{\xi}]$ are the leading-order nonlinear forces (see \citealt{Kumar:89, VanHoolst:94, Schenk:02}).  Following \citet{Schenk:02}, we use the linear eigenmodes to expand the six-dimensional phase space vector  
\begin{eqnarray}
\label{eq:xiexpansion}
 \left[ \begin{array}{c}
        \gv{\xi}(\gv{x},t)\\
        \dot{\gv{\xi}}(\gv{x},t)
        \end{array} \right] =
\sum_a q_a(t) \left[ \begin{array}{c}
        \gv{\xi}_a(\gv{x}) \\
        - i \omega_a \gv{\xi}_a(\gv{x})
        \end{array} \right],
\end{eqnarray}
where each eigenmode is specified by its amplitude $q_a$, frequency $\omega_a$, and eigenfunction $\gv{\xi}_a(\gv{x})$. The sum over $a$ runs over all mode quantum numbers (radial order $n_a$, angular degree $\ell_a$, and azimuthal order $m_a$) and frequency signs to allow both a mode and its complex conjugate. We normalize the eigenfunctions such that  $2 \omega_a^2\int d^3x \rho \, |\gv{\xi}_a|^2 = E_\ast$, where $\rho$ is the density and $E_\ast \equiv GM^2/R$ is a characteristic energy of a star of mass $M$ and radius $R$. The energy of a mode is therefore given by $E_a=|q_a|^2 E_\ast$; throughout the paper we will express energies in units of $E_\ast$. Plugging the expansion into Equation (\ref{eq:xieqn}) and using the orthogonality of eigenmodes leads to a set of coupled nonlinear amplitude equations for each oscillation mode 
\bea
\dot{q}_a + \left(i\omega_a +\gamma_a\right) q_a & =& i\omega_a f_a(t)+ i \omega_a \sum_{bc} \kappa_{abc}^* q_b^* q_c^*
\label{eq:modeampeqn},
\eea
where $\gamma_a$ is the linear damping rate of mode $a$, $f_a$ is its linear driving, $\kappa_{abc}$ is the dimensionless three-mode coupling coefficient, and asterisks denote complex conjugation.

We use the \texttt{MESA} stellar evolution code \citep{Paxton:11, Paxton:13, Paxton:15, Paxton:18} to construct RGB models with mass $M=\{1.2, 2.0, 2.5\} M_\odot$ and $\nu_{\rm max}\simeq \{50,100, 150, 200\}\muHz$, which coincide with the range observed by \emph{CoRoT} and \emph{Kepler}. We use the \texttt{GYRE} stellar oscillation code \citep{Townsend:13, Townsend:18} to find the eigenmodes of each stellar model; we use the adiabatic solutions to calculate the mode frequencies and $\kappa_{abc}$ and the non-adiabatic solutions to calculate the radiative damping rates (see below; since $\gamma_a \ll \omega_a$ for the modes we consider, the modes are nearly adiabatic). To integrate the amplitude equations, we use the \texttt{CVODES} Adams solver from the \texttt{SUNDIALS} package \citep{Hindmarsh:05} and parallelize the computations across multiple CPUs using standard parallelization techniques.

The calculations of $\gamma_a$ and $f_a$ are described in  Sections~\ref{sec:linear_damping} and ~\ref{sec:stochastic_driving} below, respectively.  The coupling coefficient $\kappa_{abc}=E_\ast^{-1}\int d^3 x\, \gv{\xi}_a\cdot \gv{f}_2[\gv{\xi}_b, \gv{\xi}_c]$.  It is symmetric in the three indices and we calculate it using the expressions given in {\protect\NoHyper\citeauthor{Weinberg:12} \protect\endNoHyper}(\citeyear{Weinberg:12}; see their A55-A62). Its dependence on mode parameters and the stellar model are described in detail in \citetalias{Weinberg:19}.  

\begin{figure}
\centering
\includegraphics[width=3.6in]{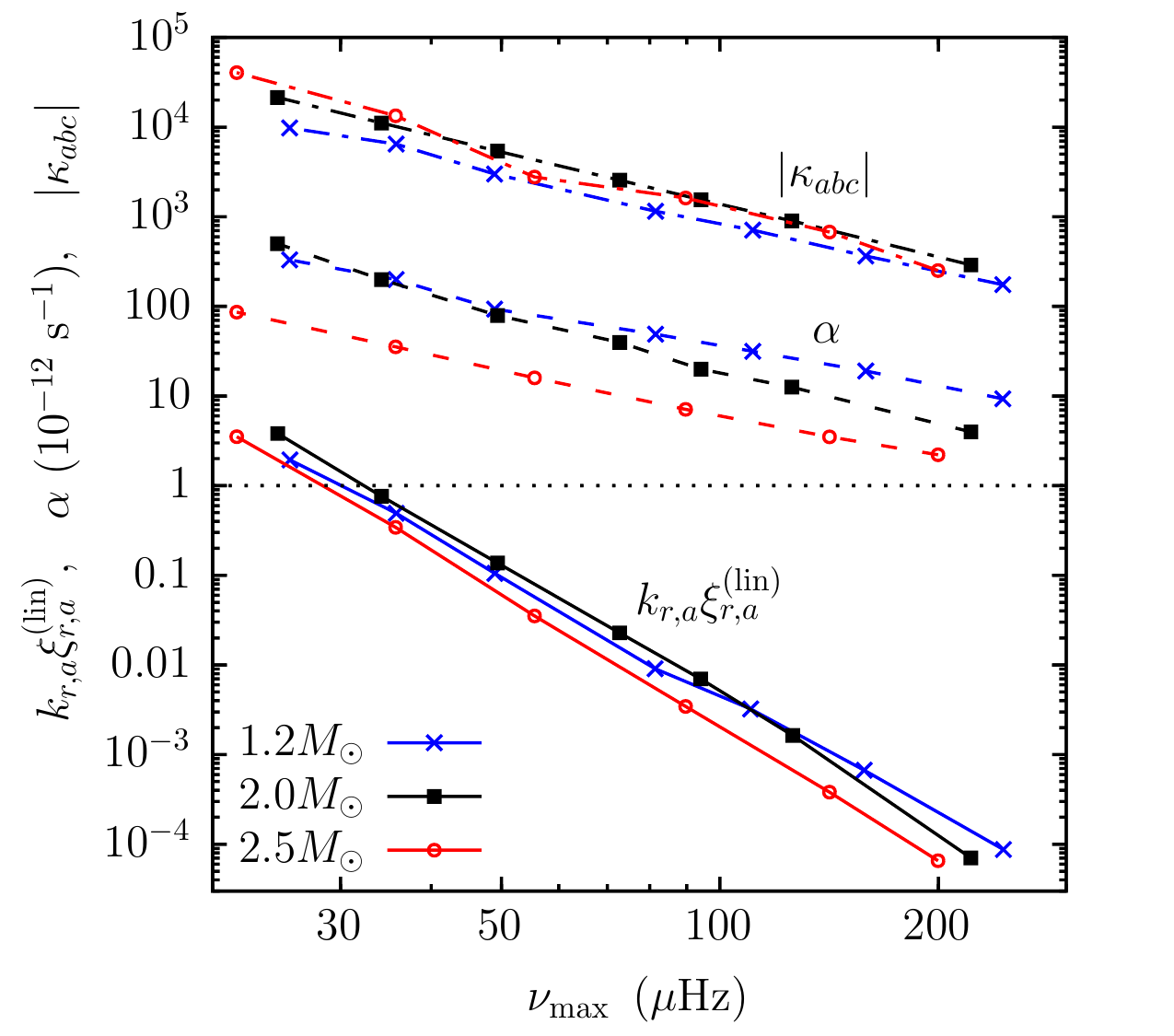} 
\caption{
Nonlinear coupling coefficient $\kappa_{abc}$ (dashed-dotted lines), and linear damping coefficient $\alpha$ (dashed lines; in units of $10^{-12}\trm{ s}^{-1}$) as a function of $\nu_{\rm max}$ for $M=1.2 M_\odot$ (blue curves with crosses), $2.0 M_\odot$ (black curves with squares), and $2.5 M_\odot$ (red curves with circles). The modes  are $(\ell_a, \ell_b,\ell_c)=(1,2,3)$ with $\nu_a\simeq \nu_{\rm max}$ and $\nu_b+\nu_c\simeq \nu_a$.  The solid lines show the maximum shear $k_{r,a}\xi_{r,a}^{(\rm lin)}$ of $p$-$m$ mixed-modes at their linear energy (Eq.~\ref{eq:Ea_obs}).
\label{fig:gamma_kappa}}
\end{figure}

As shown in Figure 1 of \citetalias{Weinberg:19}, the eigenfunctions of high-order mixed modes steepen in the core of a red giant due to geometric focusing.  As a result, the nonlinearities are greatest near the center of the star, which is therefore where the nonlinear mode coupling and contributions to $\kappa_{abc}$ peak. In Figure~\ref{fig:gamma_kappa} we show  the maximum $|\kappa_{abc}|$ as a function of the frequency of maximum power $\nu_{\rm max}$ assuming $\nu_a=\omega_a/2\pi\simeq \nu_{\rm max}$, resonant daughters, and $(\ell_a,\ell_b,\ell_c)=(1,2,3)$.  As explained in \citetalias{Weinberg:19}, the coupling coefficient scales as $\kappa_{abc}\simeq \kappa_0 (\nu_a/100\muHz)^{-2}$.  We find $\kappa_0\simeq \{900, 1400, 1400\}$ for $M=\{1.2, 2.0, 2.5\} M_\odot$.  

Figure~\ref{fig:gamma_kappa} also shows the maximum shear $k_{r, a} \xi_{r,a}^{(\rm lin)}$ of dipole $p$-$m$ modes (convective envelope-trapped mixed modes) at linear amplitude, where $k_{r,a}$ is the radial wavenumber and $\xi_{r,a}^{(\rm lin)}=\xi_{r,a} (E_{a, \rm lin}/E_\star)^{1/2}$ is the radial displacement at linear energy (see Eq.~\ref{eq:Ea_obs}). If $k_{r,a}\xi_{r,a}^{(\rm lin)}\ga 1$, the perturbation is strongly nonlinear and likely to overturn and break before actually reaching linear amplitude.  We see that the modes are only weakly nonlinear for $\numax \ga 50\muHz$, suggesting that accounting for only the lowest-order nonlinearities (three-mode coupling) is sufficient.
 
\subsection{Linear damping rate $\gamma_a$} 
\label{sec:linear_damping}
The total linear damping rate $\gamma_a = \gamma_{a, {\rm rad}}+\gamma_{a, {\rm conv}}$ has contributions from the heat radiated away each oscillation cycle ($\gamma_{a, {\rm rad}}$) and from interactions between the mode and convective motions in the envelope ($\gamma_{a, {\rm conv}}$). To determine the contribution from radiative damping, we solve the non-adiabatic oscillation equations using \texttt{GYRE}.   We find (see also \citetalias{Weinberg:19})
\beq
\gamma_{a, {\rm rad}} \simeq \alpha \Lambda_a^2 \left(\frac{\nu_a}{100\muHz}\right)^{-2},
\eeq
where $\Lambda_a^2 = \ell_a(\ell_a+1)$ and $\alpha(M,\nu_{\rm max})$ is a model dependent constant.  This damping is dominated by the radiative core, where the mixed mode wavelength is short. Values of $\alpha$ are shown in Figure~\ref{fig:gamma_kappa}. It increases with decreasing $\nu_{\rm max}$ and is in the range $\alpha\simeq [0.01, 10]\times10^{-10}\trm{ s}^{-1}$ for $M=[1.2, 2.5]M_\odot$ and $\nu_{\rm max}=[30, 200]\muHz$. 

Convection's contribution to mode damping is not well understood but has been modeled using non-local, time-dependent treatments that consider the variations in the convective flux and  turbulent pressure due to an oscillation mode  (see \citealt{Dupret:09} and \citealt{Grosjean:14} and references therein).  Given the models' uncertainties and mismatch with observations (see below), we estimate $\gamma_{a, {\rm conv}}$ from observables using the approach described in \citetalias{Weinberg:19}.  In particular, following arguments made by \citet{Dupret:09} and \citet{Grosjean:14} (see also \citealt{Benomar:14}),  we use the fact that the low-$\ell$ modes we consider have a structure in the convective envelope that is similar to that of the radial modes.  As a result,  $\mathcal{P}_0 \mathcal{M}_0 \simeq \mathcal{P}_a \mathcal{M}_a$, where $\mathcal{P}_a$ is the time-averaged power
supplied to the mode by turbulent convection, $\mathcal{M}_a=M I_a$ is the mode mass, $I_a$ is the mode inertia, and $0$ subscripts denote radial mode quantities.  For the linear problem, the mode energy $E_a=\mathcal{P}_a/2\gamma_a$ and if we assume the modes all have nearly the same linear energy (i.e., $E_a \simeq E_0$), it follows that $\gamma_{a,\rm conv} \simeq \gamma_0 \mathcal{M}_0/\mathcal{M}_a$.  Thus, given $\gamma_0$ from observations, as described in the next paragraph, and the calculation of mode masses using \texttt{GYRE}, we obtain $\gamma_{a,\rm conv}$.  

\citet{Vrard:18} fit the frequency spectra of more than five-thousand  \emph{Kepler} red giants to determine the resolved linewidths $\Gamma_0 =\gamma_0/\pi$ of radial modes with $\nu_0\simeq \nu_{\rm max}$.  They found $\Gamma_0(\nu_{\rm max})\simeq [0.05, 0.2] \muHz$ (lifetimes $\tau_0 = \gamma_0^{-1} \simeq 20-70\trm{ days}$) over the range $M\simeq[0.8, 2.5]M_\odot$ and $\nu_{\rm max}\simeq [10, 200]\muHz$.  By contrast, theoretical calculations \citep{Dupret:09, Grosjean:14} predict lifetimes of $\approx 2-10\trm{ days}$, which is significantly shorter than the measured lifetimes.  That the dissipation is much less efficient than the models predict is indicative of our incomplete understanding of convective damping.

For $p$-$m$ modes, $\mathcal{M}_a / \mathcal{M}_0 \approx 3-5$ and thus their lifetimes should be longer than the radial modes by this factor if $\gamma_{a, {\rm conv}}\simeq \gamma_0 \mathcal{M}_0/\mathcal{M}_a$   (consistent with the {\it ratio} of lifetimes shown in \citealt{Dupret:09} and \citealt{Grosjean:14}). Since $\tau_0 = \gamma_0^{-1} \simeq 20-70\trm{ days}$, this implies that for $p$-$m$ modes  $\tau_a \simeq 50 - 400\trm{ days}$  ($\gamma_a \simeq [3-20]\times10^{-8}\trm{ s}^{-1}$).  To examine how our results depend on $\gamma_a$, we will consider  $p$-$m$ modes with  $\gamma_a=10^{-7}\trm{ s}^{-1}$ and $\gamma_a=10^{-8}\trm{ s}^{-1}$ in our mode network calculations (see Section~\ref{sec:parents}). Given that the observations imply $\gamma_a \simeq [3-20]\times10^{-8}\trm{ s}^{-1}$, a damping rate of  $\gamma_a=10^{-7}\trm{ s}^{-1}$  might be considered typical while  $\gamma_a=10^{-8}\trm{ s}^{-1}$ would be on the low end of the likely values.

\begin{figure}
\centering
\includegraphics[width=3.6in]{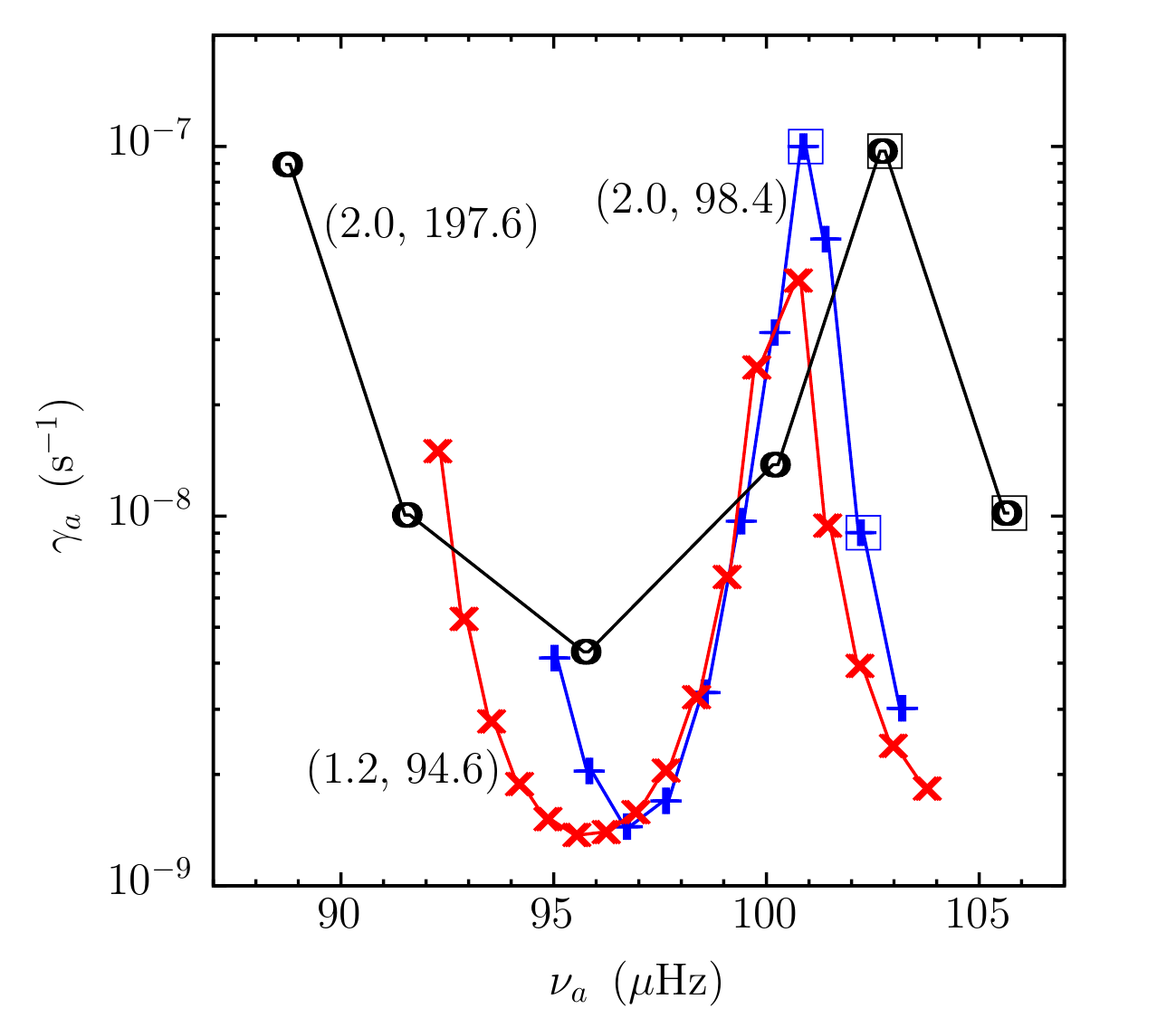} 
\caption{
Linear damping rate of the $\ell_a=1$ parent modes from three of our networks as a function of mode frequency.  Here we set $\gamma_{a,\rm conv}$ such that the $p$-$m$ modes have a damping rate  $\gamma_p\simeq 10^{-7}\trm{ s}^{-1}$.  The labels indicate each model's $(M/M_\odot, \nu_{\rm max}/\mu\trm{Hz}$).  For clarity, we connect the points with straight lines and shift the black points ($\nu_{\rm max}=197.6\muHz$)  to the left by $100\muHz$.   The two blue squares (black squares) mark the $m_a=1$, $n_a=93, 95$ ($n_a=26,27$) parent modes shown in Figure~\ref{fig:Eth_parent}.
\label{fig:gamma_parents}}
\end{figure}

In Figure~\ref{fig:gamma_parents} we show $\gamma_a$ as a function of $\nu_a$ for the $\ell_a=1$ parent modes from three of our networks (see Section~\ref{sec:networks}).  The modes are consecutive in $n_a$ and all lie near $\numax$. The $n_a$ range is chosen to span a half cycle between acoustic peaks. Notice that the $p$-$m$ modes have significantly larger damping rates  than the $g$-$m$ modes (similar to \citealt{Dupret:09, Grosjean:14}). This is because the $p$-$m$ modes have much smaller mode inertias  and thus convective motions in the envelope are more effective at damping them. Importantly, by considering parents that span the range from $p$-$m$ to $g$-$m$ modes, we will be able to study how linear damping influences the nonlinear equilibria of modes.

\subsection{Stochastic driving $f_a(t)$}
\label{sec:stochastic_driving}

The linear driving term $f_a(t)$ in Equation (\ref{eq:modeampeqn}) is designed to model the stochastic excitation of mode $a$ by turbulent motions at the top of the convective envelope.  As in \citetalias{Weinberg:19} (see also, e.g., \citealt{Kumar:88, Chang:98}), we treat the forcing as a Poisson process involving a random sequence of  impulses at times $t_j$.  We assume that the time between consecutive  impulses, $\Delta t=t_{j+1}-t_j$, is an independent random variable whose probability density is given by $p(\Delta t) = \mu \exp(-\mu \Delta t)$, where $\mu$ is the mean number of impulses per unit time.  The mode forcing is the sum of all the individual impulses
\beq
f_a(t) = f_{0,a} \sum_j c_j \psi_j(t),
\eeq
where we assume that the duration of each impulse has a Gaussian time dependence  $\psi_j(t)=\exp[-(t-t_j)^2/2\tau_{\rm cor}^2]$ with a correlation time $\tau_{\rm cor}$.  Similar to the model by \citet{Chang:98}, we assume that the complex amplitude  $c_j = (E_j/\bar{E}_a)^\alpha  Y_a(\theta_j,\phi_j)$, where $E_j$ and $\alpha$ are real numbers and $Y_a=Y_{\ell_a m_a}$ is the spherical harmonic function representing mode $a$.   We take $\alpha=1$ and  sample $E_j$ from an exponential distribution with mean $\bar{E}_a$ to give a Boltzmann distribution of energies.\footnote{In the model of \citet{Chang:98}, $E_j$ represents the random energy of the convective eddies whose size distribution determines $\alpha$. The latter is uncertain, with $3/4 \lesssim \alpha \lesssim 3$ possible. While we present results assuming $\alpha=1$, we carried out numerical  experiments with $\alpha=3$ and found similar results.}  We set $\bar{E}_a=E_{a, \rm lin}$, where $E_{a, \rm lin}$ is the linear energy of the mode as inferred from observations (see Section~\ref{sec:forcing_params}). The angular variables  $\cos\theta_j$ and $\phi_j$ in $Y_a(\theta_j,\phi_j)$ are sampled from flat distributions assuming the parents are $\ell_a=|m_a|=1$ modes with $\cos\theta_j$ in the range $[0, 1]$ (since here $\theta_j$ is symmetric about $\pi/2$) and $\phi_j$ in the range $[0, 2\pi]$.  This mimics the angular distribution of the eddies, which are treated as points since their sizes are much smaller than $R$. Note that since $f_a(t)$ is a smoothly varying function of time and we resolve the timescales of the individual impulses, we do not need to use a solver dedicated to stochastic differential equations.  We have confirmed that our numerical solution of the linear problem converges on the expected solution, i.e., the power at frequency $\omega$ is a Lorentzian with damping $\gamma_a$ and detuning $\omega-\omega_a$ (see, e.g., \citealt{CD:89}).

The observed distribution of mode energies can be fit by a Gaussian envelope with an FWHM of $\delta \nu_{\rm env} \simeq 0.66(\nu_{\rm max}/\trm{$\mu$Hz})^{0.88}\muHz$ \citep{Mosser:12:depressed}.  The resonant daughter modes in our network, which have frequencies near $\nu_a/2\simeq \nu_{\rm max}/2$, therefore have linear energies that are $\lesssim 1\%$ that of the parents'; the linear energy of the later generation modes are smaller yet.  We therefore only account for linear driving of the parent modes in our networks. As we explain in Appendix~\ref{sec:coord_xfm},  this simplification enables us to take advantage of a coordinate transformation that significantly speeds up the numerical integrations.

\subsubsection{Values of the linear forcing parameters}
\label{sec:forcing_params}

Our model for stochastic forcing $f_a(t)$ depends on three parameters: $\tau_{\rm cor}$, $\mu$, and $f_{0,a}$. We determine them as follows.

The correlation time $\tau_{\rm cor}$ is expected to be on the order of the eddy turnover time $\tau_{\rm eddy}$.   Since modes with frequency near $\tau_{\rm eddy}^{-1}$ are most strongly excited and thus lie near the peak of the power spectrum \citep{Goldreich:88},  we set $\tau_{\rm cor}=(2\pi \nu_{\rm max})^{-1}$.   Experiments in which we increased or decreased $\tau_{\rm cor}$ by factors of a few relative to this (while keeping the parent energy fixed by also varying $f_{0,a}$) yielded similar results.
 
Since the size of each granule is of order a scale height $H$, there are approximately $(R/H)^2 \gg 1$ granules, and the mean time between impulses $\Delta t \sim \tau_{\rm eddy} (H/R)^2 \ll \omega_a^{-1}$. While this suggests $\mu \gg 1$ per mode period, we find that our numerical results are insensitive to $\mu$ as long as we set $\mu \ga \gamma_a$, i.e., several impulses per mode damping time.  Since the numerical integrations run   faster for smaller $\mu$, we set $\mu = 0.01 \nu_{\rm max}$; experiments with higher $\mu$ yielded similar results.

For our model of stochastic driving, the ensemble average of the parent linear energy is given approximately by \citepalias{Weinberg:19}
\beq
E_{a, \rm lin}/E_\ast = \langle |q_{a, \rm lin}(t)|^2  \rangle \approx (f_{0,a}\omega_a \tau_{\rm cor})^2 \left(\frac{\mu}{2\gamma_a}\right),
\eeq
where angle brackets denote an average over all realizations of the random process and $q_{a, \rm lin}(t)$ is the linear amplitude of the mode given by the solution of Equation (\ref{eq:lin_amp}).
Thus, given $\tau_{\rm cor}$ and $\mu$, we determine $f_{0,a}$ for a mode of a given $\omega_a$ and $\gamma_a$ by adjusting its value until the computed $E_{a, \rm lin}$   matches the ``observed" value.  We determine the latter using the approach described in \citetalias{Weinberg:19}, which  combines the measured linewidths of radial modes $\Gamma_0$ (see Section~\ref{sec:linear_damping}) with calculations of the power $\mathcal{P}_0$ supplied to radial modes by turbulent convection.  According to the 3D hydrodynamics models of \citet{Samadi:11}, the power scales strongly with a star's luminosity-to-mass ratio $L/M$, with $\mathcal{P}_0(M,\nu_{\rm max})=B x^s$, where $x = (L/L_\odot)(M_\odot/M)$, $B=4.2^{+1.0}_{-0.8}\times 10^{22} \trm{ erg s}^{-1}$, and $s=2.60\pm 0.08$.  Since low-$\ell$ modes near $\nu_{\rm max}$ are expected to all have similar linear energies \citep{Dupret:09,Grosjean:14},  $E_{a, \rm lin} \simeq \mathcal{P}_0 /2\pi\Gamma_0$ and thus
\bea
&E_{a, \rm lin}(M, \nu_{\rm max})&\simeq \left(1.8^{+0.4}_{-0.3}\right)\times10^{-16} E_\ast
\non &&
\hspace{-1.8cm}
\times  \left(\frac{\beta}{1.5}\right)^2 \left(\frac{\Gamma_0}{0.1\muHz}\right)^{-1}
\left(\frac{M}{1.5M_\odot}\right)^{-3/2}
\non &&
\hspace{-1.8cm}
\times\left(\frac{T_{\rm eff}}{4800\trm{ K}}\right)^{8.9\pm0.3}\left(\frac{ \nu_{\rm max}}{100\muHz}\right)^{-3.1\mp 0.1},
\label{eq:Ea_obs}
\eea
where $T_{\rm eff}$ is the effective temperature and we used the scaling relations $L\propto R^2 T_{\rm eff}^4$ and $\nu_{\rm max}\propto M R^{-2}T_{\rm eff}^{-1/2}$  with solar reference values $\nu_{\rm max,\odot}= 3101\muHz$ and $T_{\rm eff,\odot}=5777\trm{ K}$ \citep{Kjeldsen:95, Stello:09, Huber:10}. The uncertainties in the expression arise from the uncertainties in \citeauthor{Samadi:11}'s calculation of $\mathcal{P}_0$.   The  constant $\beta\approx 1.5$, introduced by \citet{Samadi:12}, corrects for an overall offset between the observed bolometric amplitudes and those predicted by their hydrodynamical models.

\section{\bf M\lowercase{ode} N\lowercase{etworks}}
\label{sec:networks}

In this Section, we describe our procedure for constructing the 24 mode networks whose amplitude equations we integrate. We begin in Section~\ref{sec:parametric} with a discussion of the stochastic parametric instability and the key parameters that determine whether a mode is unstable. In Section~\ref{sec:constructing}, we describe how we select each generation of modes that comprise a network, from parents to daughters, to  granddaughters, etc., and present the detailed structure of a representative network.

\subsection{Parametric instability of parent modes}
\label{sec:parametric}
At sufficiently large amplitude, a mode can excite resonant secondary modes via the  parametric instability.  The stability criterion for a stochastically driven parent mode is found by analyzing the stochastic Mathieu equation  (see, e.g., \citealt{Stratonovich:65, Ariaratnam:76, vanKampen:92, Zhang:93, Poulin:08}).  In \citetalias{Weinberg:19}, we studied the stability of modes in red giants consisting of a single stochastically driven parent mode coupled to a pair of resonant daughter modes.  We showed that a parent mode $a$  excites a pair of daughter modes $b,c$ if the parent's linear energy $E_{a, \rm lin}$ exceeds a threshold energy 
\bea
E_{a, \rm th}
&\simeq& \frac{\gamma_a \sqrt{\gamma_b\gamma_c}}{\kappa_{abc}^2\omega_b\omega_c}\left[1+\frac{\Delta_{bc}^2}{\gamma_a^2}\right]\, E_\star \non
 &\simeq &2.5 \times 10^{-16}\, E_\star \left(\frac{\gamma_a}{10^{-7} \trm{ s}^{-1}}\right)\left(\frac{\gamma_b}{10^{-9} \trm{ s}^{-1}}\right)
\non &&\times\left(\frac{\kappa_{abc}}{10^3}\right)^{-2} 
\left(\frac{\omega_b/2\pi}{100 \muHz}\right)^{-2}
\left[1+\frac{\Delta_{bc}^2}{\gamma_a^2}\right],
\label{eq:Eth}
\eea
 where the expression assumes $\gamma_a \gg \gamma_{b}\simeq \gamma_c$ and $|\Delta_{bc}| \ll \omega_a$.  
By comparing $E_{a, \rm th}$ with $E_{a, \rm lin}$ (Equation~\ref{eq:Ea_obs}), and also by numerically integrating the three-mode amplitude equations, \citetalias{Weinberg:19} found that dipole modes near the power maximum  ($\nu_a\simeq \numax$) can be parametrically unstable over a broad range in red giant mass and evolutionary state ($M\gtrsim 1.0 M_\odot$ and $\numax \lesssim 200\muHz$; see Figure~4 in \citetalias{Weinberg:19}).  

\begin{figure}
\centering
\includegraphics[width=3.4in]{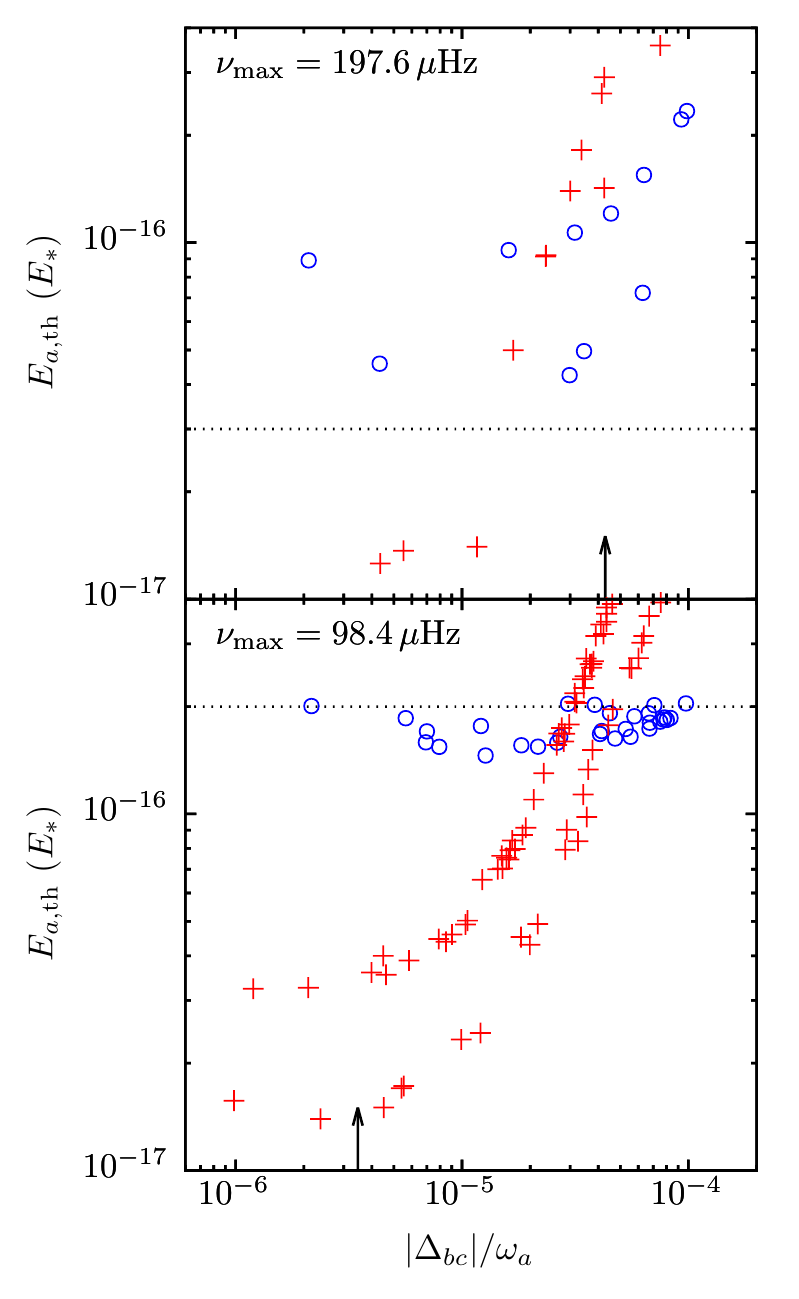} 
\caption{
Nonlinear energy threshold $E_{a, \rm th}$ of  parent modes $a$ coupled to daughter pairs  $b,c$ as a function of their detuning $\Delta_{bc}=\omega_a + \omega_b+\omega_c$.  The modes are from the networks of the $M=2.0 M_\odot$ models with $\nu_{\rm max}\simeq197.6\muHz$ (top panel) and $\nu_{\rm max}\simeq 98.4\muHz$ (bottom panel).  The red crosses and blue circles correspond, respectively, to the $\gamma_a\simeq 10^{-8}\trm {s}^{-1}$ and $\gamma_a\simeq 10^{-7}\trm {s}^{-1}$ parent modes marked with boxes in Figure~\ref{fig:gamma_parents}.   The daughters are the lowest threshold pairs with $l_b \le 2$ and $l_c=l_b+1$. The horizontal dotted lines show the estimated linear energy $E_{a, \rm lin}$ of the parent modes (Eq.~\ref{eq:Ea_obs}).  The vertical arrows show the analytic estimate of the minimum detuning $|\Delta_{bc}|/\omega_a \approx (4\ell_b^3 n_a^2)^{-1}$ \citep{Wu:01} assuming $\ell_b=2$.  
\label{fig:Eth_parent}}
\end{figure}

At a given $\nu_a$, $E_{a,\rm th}$ depends on four parameters ($\gamma_a$, $\gamma_b$, $\Delta_{bc}$, $\kappa_{abc}$). However, in practice we find that the value of $\gamma_a$, which can vary significantly from parent-to-parent (see Figure~\ref{fig:gamma_parents}), often determines whether a particular parent mode is stable.  We illustrate this point in Figure~\ref{fig:Eth_parent}, which shows $E_{a, \rm th}$ as a function of $\Delta_{bc}$ for parent and daughter modes from two of our $M=2.0M_\odot$ models.  As the parent damping rate increases from $\gamma_a \simeq 10^{-8}\trm{ s}^{-1}$ (red crosses) to $10^{-7}\trm{ s}^{-1}$ (blue circles), the minimum $E_{a, \rm th}$ goes from well below $E_{a, \rm lin}$ (dotted horizontal lines) to close to or greater than $E_{a, \rm lin}$.  Since most $\ell_a=1$ modes near $\numax$ have $\gamma_a \ll 10^{-7}\trm{ s}^{-1}$ (see Figure~\ref{fig:gamma_parents}), most parent modes will be unstable. However, the $p$-$m$ modes can be stable if $\Gamma_0 \gtrsim 0.1 \muHz$ since they then have $\gamma_a \gtrsim 10^{-7}\trm{ s}^{-1}$ (Section~\ref{sec:linear_damping}).  Interestingly, the  observations span $\Gamma_0\approx 0.05-0.2\muHz $ for otherwise similar stars \citep{Vrard:18}.  This suggests that the stability of $p$-$m$ modes depends on where the particular star under consideration happens to fall in the distribution of $\Gamma_0$ values.\footnote{When searching for daughter pairs that minimize $E_{a, \rm th}$,  we find that there are always pairs with small enough $|\Delta_{bc}|$ to ensure $E_{a, \rm th} < E_{a, \rm lin}$ provided $\gamma_a \lesssim 10^{-7}\trm{ s}^{-1}$. \citet{Wu:01} argue that, to 
an order-of-magnitude, the minimum detuning  $|\Delta_{bc}|/\omega_a \approx (4\ell_b^3 n_a^2)^{-1}$. In Figure~\ref{fig:Eth_parent}  we see that their expression is reasonably accurate, although it tends to overestimate the results of numerical searches by factors of a few to ten.}

\subsection{Constructing mode networks}
\label{sec:constructing}

\citetalias{Weinberg:19} only study small sets of modes consisting of parent-daughter triplets. Although this provides information about where in the $M$-$\numax$ plane modes are unstable, in order to determine how their instability impacts observables (e.g., oscillation amplitudes), we need to know how the instability saturates, i.e., the nonlinear equilibrium of the modes. We find that the unstable parents drive the daughters to such large amplitudes that the daughters themselves parametrically excite  granddaughters and then the granddaughters excite great-granddaughters etc..  Thus, solving for the saturation entails integrating a large network of coupled modes across several generations.  One of the challenges when constructing mode networks is figuring out which modes to include, and how many generations, in order to ensure that the system converges on the true nonlinear equilibrium.  

We describe how we select the parents and later generation modes in the next two subsections.  We note first that the mode frequencies in our networks are all rotationally split assuming a uniformly rotating star with spin frequency $\Omega_{\rm s}$ (we otherwise ignore the impact of rotation on the eigenmodes).  We compute the mode frequencies as
\beq
\omega_{a} = {\omega}_{n_a, \ell_a, m_a=0} + m_a\lambda_a \Omega_{\rm s},
\eeq
where  $\lambda_a$ is a mode-dependent constant (equal to one minus the Ledoux constant).  We find that the network structure and integration are relatively  insensitive to the particular magnitude of  $m_a \lambda_a \Omega_{\rm s}$.  That is because the principal effect of including rotational splitting is simply to decrease the minimum detuning by increasing the number of modes with distinct frequencies (since, as we show below, the typical detuning is far smaller than $m_a \lambda_a \Omega_{\rm s}$).  Specifically, rotational splitting decreases $|\Delta_{bc}|$ by a factor of order $\ell_b$ \citep{Wu:01} and since $\ell_b\le 3$ for all the modes of our networks, it is not an especially large effect.   Given this insensitivity, for simplicity  we assume a spin period $P_{\rm s}=2\pi/\Omega_{\rm s}=100\trm{ day}$ in all our networks. This choice of $P_{\rm s}$ is motivated by asteroseismic measurements of the core and envelope spin frequency in stars on the red giant branch, which range from $\sim 0.1$ to $1\muHz$ (see, e.g., \citealt{Deheuvels:12, Deheuvels:14, Mosser:12:rotation} and Figure~4 in the the review by \citealt{Aerts:19} and references therein).  We assume $\lambda_a = 1 - \left[\ell_a(\ell_a+1)\right]^{-2}$, which is  appropriate for high-order $g$-modes \citep{Aerts:10}.  A more accurate method would be to use \texttt{GYRE}'s calculation of $\lambda_a$ for each mode; however, this slows the mode search considerably and since the results are not sensitive to the magnitude of the splitting, we find that the analytic estimate is sufficient for our purposes. 

\begin{deluxetable}{CCCRCC}
\tablehead{
\dcolhead{M} & \dcolhead{R} & \dcolhead{T_{\rm eff}}& \dcolhead{\nu_{\rm max}}&\dcolhead{E_{a, \rm lin}}  & \dcolhead{n_a} \\[-1.5ex] 
\dcolhead{(M_\odot)} & \dcolhead{(R_\odot)} & \dcolhead{(\trm{K})}& \dcolhead{(\mu\trm{Hz})} &\dcolhead{(10^{-16} E_\star)} & \dcolhead{[\trm{min},\trm{max}]} 
}
\startdata
1.2& 4.6 & 4767  & 195.0&  0.5 & [44,49]   \\
1.2& 5.1 & 4735 & 158.8&  1.0 & [64,73]   \\
1.2& 6.6 & 4645 &  94.6&  4.0 & [116,133]   \\
1.2& 9.2 & 4513 &  48.9& 30.0 & [286,310]  \\
\hline
2.0& 5.8 & 4964 & 197.6& 0.3 & [26,31]    \\
2.0& 6.7 & 4893 & 147.4& 0.8 & [45,55]   \\
2.0& 8.2 & 4802 & 98.4& 2.0 & [92,102]   \\
2.0& 11.7& 4639&  49.4& 10.0 & [267,281] \\
\hline
2.5& 6.0 & 6872 & 195.9& 3.0 & [3,6]    \\
2.5& 7.4 & 5188 & 147.3& 0.7 & [26,34]   \\
2.5& 9.1 & 5074 &  98.8& 2.0 & [56,69]   \\
2.5& 13.0& 4922 & 49.2& 10.0 & [175,195] \\
\enddata
\caption{Parameters of the stellar models and parent modes of each network. The linear energy of the parent modes $E_{a,\rm lin}$ is given by Equation~(\ref{eq:Ea_obs}). \label{tab:parameters}}
\end{deluxetable}


\subsubsection{Parent modes}
\label{sec:parents}
Given a stellar model, the first step in constructing a mode network is to select the parent modes.  We assume the parents are $\ell_a=1$ modes with $\nu_a \simeq \numax$, which we find using \texttt{GYRE}.  We focus on these modes because they are the mixed modes best resolved by \emph{Kepler} and because we want to know if nonlinear damping can explain why their amplitude is suppressed in some stars.  Moreover, dipole modes are more likely than radial modes to be parametrically unstable because the latter have smaller amplitudes in the core and thus smaller $\kappa_{abc}$. The parents in the networks are spaced consecutively in radial order $n_a$ and have azimuthal order $m_a=-1,0,1$.  We choose the range in $n_a$ such that the parents span a full cycle from convective envelope-trapped ($p$-$m$) modes to  radiative core-trapped ($g$-$m$) modes.  We center the $n_a$ range such that it includes the $p$-$m$  mode closest to $\numax$.   Figure~\ref{fig:gamma_parents} shows the span of parent modes used in three of our networks.   
  
Table~\ref{tab:parameters} gives the range in $n_a$ and the values of $E_{a,\rm lin}$ for each of our networks. In order to account for the observed range in $\Gamma_0$ and the theoretical uncertainty  in the convective contribution to parent mode damping, we consider two values for the maximum of $\gamma_{a, \rm conv} \approx \pi \Gamma_0 \mathcal{M}_0/\mathcal{M}_a$.  The values are chosen such that $\gamma_p$, the damping rate of the most $p$-$m$-like mode of a network (i.e., the mode with the smallest mode inertia $\mathcal{M}_a$), is either  $\gamma_p\simeq\gamma_{a, \rm conv}\simeq 10^{-8}\trm{ s}^{-1}$ or $\gamma_p\simeq 10^{-7}\trm{ s}^{-1}$ (e.g., Figure~\ref{fig:gamma_parents} assumes the latter). We name the networks according to their $M$, $\numax$ and  $\gamma_p$.  For example, $M2.0\nu98.4\gamma7.0$ corresponds to the network with $M=2.0 M_\odot$, $\numax =98.4\muHz$, and $\gamma_p \simeq 10^{-7}\trm{ s}^{-1}$.  

\begin{figure}
\centering
\includegraphics[width=3.5in]{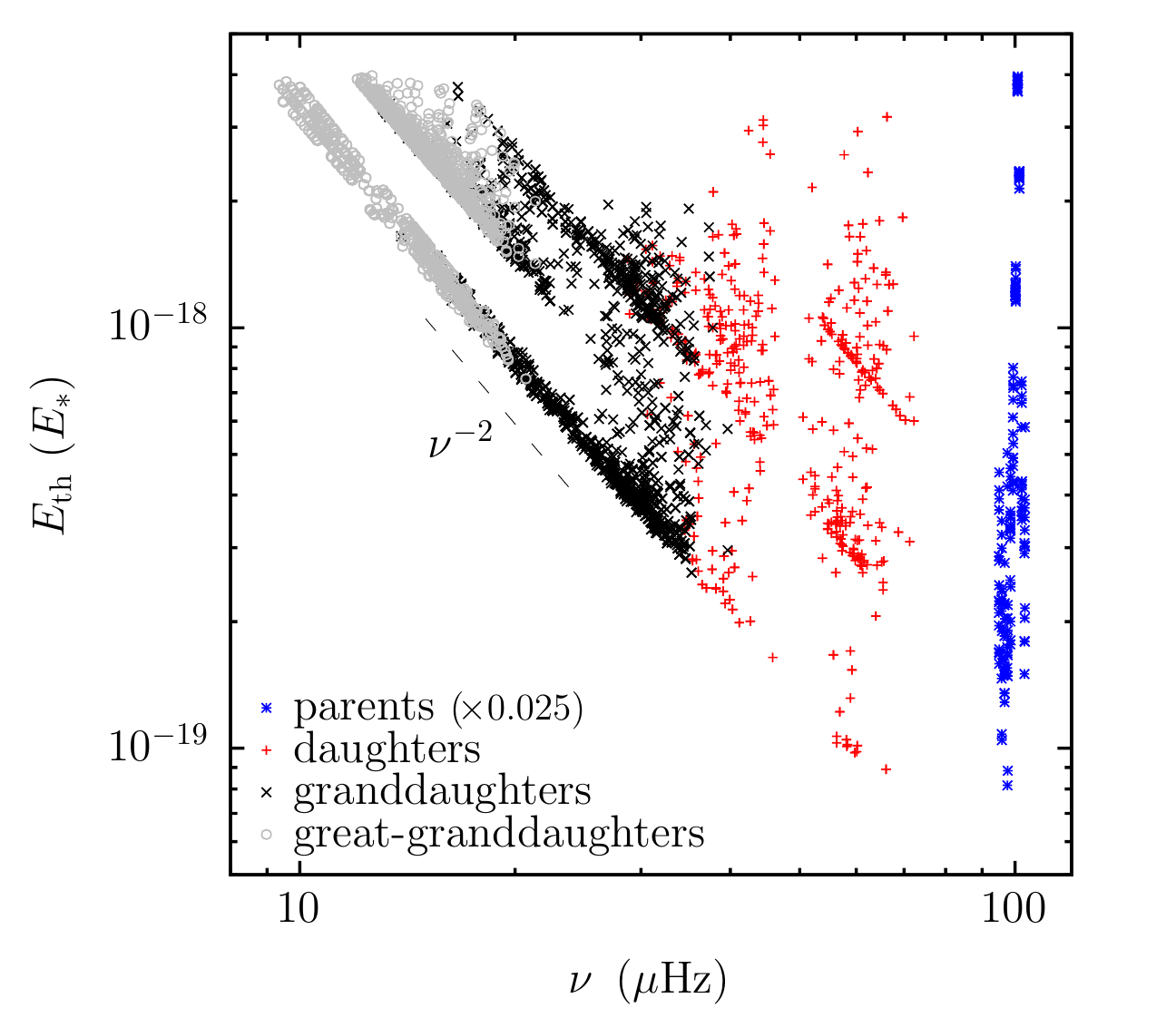} 
\caption{
Nonlinear energy threshold $E_{\rm th}$ as a function of mode frequency $\nu$ for the modes of the $M2.0\nu98.4\gamma7.0$ network.    Each generation is shown with a different colored symbol.  For display purposes, we plot the parents' $E_{\rm th}$  multiplied by $0.025$. See the text for a description of the network's structure.  
\label{fig:Eth_all_modes}}
\end{figure}

In Figure~\ref{fig:Eth_all_modes} we show $E_{\rm th}$ as a function of mode frequency for the parent modes (blue asterisks)  of the $M2.0\nu98.4\gamma7.0$ network.  The network has 33 parents (radial orders $n_a=[92,102]$ and $m_a=[-1,1]$) and since each parent of the network is coupled to five daughter pairs (Section~\ref{sec:daughters}), there are 165 distinct values of $E_{\rm th}$.  We see that they are in the range $E_{\rm th} =[0.03, 1.6]\times10^{-16} E_\ast$. (in the figure we multiply the parent values by 0.025 in order to display them on the same scale as the later generation modes, which have considerably smaller $E_{\rm th}$); the large $E_{\rm th}$ range is due to the large range in the linear damping rate $\gamma_a$ of the parents, which increases by a factor of $\simeq 100$ going from $g$-$m$ to $p$-$m$ parent modes (see Figure~\ref{fig:gamma_parents}).

\subsubsection{Daughter modes}
\label{sec:daughters}

The daughter pairs with the lowest $E_{\rm th}$ grow the fastest and are thus the most effective at damping the parent modes (see also \citealt{Essick:16}).  For each parent $a$, we therefore search for daughter pairs $b,c$ that minimize $E_{\rm th}$.  The networks include each parent's  five lowest $E_{\rm th}$ daughter pairs; experiments in which we include additional low $E_{\rm th}$ daughter pairs yield very similar nonlinear equilibria.   We restrict the search for daughter modes to $\ell_b\le3$ and $m_b=[-\ell_b, \ell_b]$ and similarly for daughter modes $c$, subject to the  angular selection rules for three-mode coupling (\citetalias{Weinberg:19}).   When searching for low $E_{\rm th}$ daughter pairs, we consider frequencies in the range $[0.25,0.75]\omega_a$. Since $\omega_b \propto \ell_b /n_b$, increasing the maximum $\ell_b$  increases the radial range, and thus the number of mode pairs to search over.  We carried out a few searches with $\ell_b \le 5$ and found that the higher $\ell_b$ modes did not typically make the list of smallest $E_{\rm th}$.  That is because typically the $\ell_{b}\le 3$ daughters already have $\Delta_{bc}\la \gamma_a$ and thus increasing $\ell_{b}$ just increases $\gamma_b$ and thus $E_{\rm th}$  (Equation~\ref{eq:Eth}). Some of the daughter pairs of the $\numax\simeq 200\muHz$ networks do, however, favor $\ell_b > 3$ modes since at higher $\numax$, the minimum $\Delta_{bc}$ is larger and the radiative damping contribution to $\gamma_a$ is smaller.
  
We use \texttt{GYRE} to find the parent modes but not the daughter (and granddaughter etc.) modes.  Instead we use the dispersion relation for high-order $g$-modes $\omega_b= (2\pi/\Delta P_{0})\Lambda_b/n_b$ \citep{Aerts:10, Hekker:17}, where  $\Delta P_0 = 2\pi^2\left(\int N d\ln r\right)^{-1}$, and the power-law relations for $\gamma_b$ and $\kappa_{abc}$ described in Section~\ref{sec:method}.  We do not use \texttt{GYRE} here because the later generations can reach frequencies as low as $\nu\simeq 10\muHz$ (either the daughters in the case of the $\numax\simeq 50\muHz$ models or the later generation modes in the larger $\numax$ models).  Such modes have radial orders $n_b > 1000$.\footnote{The linear damping rate of such high order modes is so large that they likely damp away in less than their group travel time \citepalias{Weinberg:19}.  They should therefore really be treated as traveling waves rather than standing waves.  We discuss this issue further in Section~\ref{sec:comparing_visibilities}.}  Although in principle \texttt{GYRE} can find such high order modes, in practice we find that it yields non-sequential values of $n_b$.  We believe this may occur when the wavelength of the modes becomes shorter than the size of the occasional sharp features in $N^2$ that arise from strong gradients in composition  near the center of the stellar model. Since it is not clear how realistic these sharp features are (element diffusion will tend to smooth them out) and since  non-sequential $n_b$ can artificially limit the minimum $|\Delta_{bc}|$ and thus $E_{\rm th}$, we instead use the dispersion relation when searching for later generation modes.  As a check, we constructed a network for the $M=2.0M_\odot$, $\numax=197.6\muHz$ model that consisted of eigenmodes found entirely with \texttt{GYRE}. The network's structure and integration results were very similar to that of the network constructed using our default approach.

\subsubsection{Granddaughters, great-granddaughters, etc.}
\label{sec:daughters}

The search for later generation pairs is similar to the daughter search and  also conditioned on finding the lowest $E_{\rm th}$ pairs.  There are, however, a few differences.  Since the size of the search over radial order grows considerably with each generation, we restrict the search to pairs with frequencies in the range $[0.45,0.55]$ times the frequency of the  progenitor mode.  We include the two lowest $E_{\rm th}$ pairs per progenitor mode, rather than the five lowest as with the parent-daughter couplings.  Another difference is that rather than only consider the stochastic threshold criterion (Equation~\ref{eq:Eth}), we also consider the harmonic threshold criterion (see, e.g., \citealt{Essick:16})
\beq
E_{\rm th}^{({\rm harm})}
= \frac{\gamma_d\gamma_e}{\kappa_{bde}^2\omega_d\omega_e}\left[1+\frac{\Delta_{de}^2}{(\gamma_d+\gamma_e)^2}\right],
\label{eq:Eth_harm}
\eeq
where $b$ is the daughter mode and $d,e$ are the granddaughter modes (and similarly for yet later generation coupling).  The harmonic criterion applies if the daughter mode is oscillating harmonically at its natural frequency, rather than being driven stochastically and undergoing  random changes in phase and amplitude.  It was not clear to us which criterion is most relevant for these later generations given their weak stochastic driving. Thus,  we evaluate both types of energy thresholds and assign to each pair the smaller of the two.  In practice, we find that $E_{\rm th}$ and  $E_{\rm th}^{({\rm harm})}$ are similar to within factors of a few since the damping rates between adjacent  generations are not that different.  Experiments in which we selected pairs based on only one of the thresholds, or included more than two pairs per mode, did not noticeably affect the network integration results.

We find that networks with four generations of modes (parents, daughters, granddaughters and great-granddaughters) converge on nearly the same solution as networks with five generations.  Specifically, the average parent mode amplitudes and total dissipation do not change by more than $\approx 10\%$ when we add a fifth generation.  We will see in Section~\ref{sec:example_integration} that this is because at nonlinear equilibrium, the fourth generation modes are typically below the energy threshold to excite the fifth generation.  While most of our networks therefore have four generations, a few have five.  Since we include five daughter pairs per parent and two pairs per each subsequent generation, our four generation networks contain a total of about $10+40+160 \simeq 200$ modes per parent.   Given that there are $20-50$  parents per network (see Section~\ref{sec:parents} and Table~\ref{tab:parameters}), our typical networks contain several thousand modes.  It is never exactly 210 distinct modes per parent because individual modes can appear in more than one triplet, especially in the lower $\numax$ networks.    This is because the minimum detuning scales as $|\Delta_{bc}|\propto \omega_a^3$ \citep{Wu:01} whereas $\gamma_{a}\propto \omega_a^2$.  Thus, at low frequencies a mode is more likely to have small detunings with multiple modes.

\subsubsection{Example network}
\label{sec:example_network}
Figure~\ref{fig:Eth_all_modes} shows $E_{\rm th}$ for parent-daughter coupling through great-granddaughter-great-great-granddaughter coupling from the $M2.0\nu98.4\gamma7.0$ network (this is an example of a network that includes five generations).  The other networks have a similar distribution of $E_{\rm th}$.  We find that with the exception of parent-daughter coupling discussed in Section~\ref{sec:parents}, $E_{\rm th} \sim 10^{-19}-10^{-18} E_\ast$ for this network.  

The smallest $E_{\rm th}$ occurs for daughter-granddaughter coupling (red crosses).  This is because the radial order of the modes is high enough to ensure a small detuning but still low enough to ensure a relatively small radiative damping rate.  As the dashed line in the figure illustrates, $E_{\rm th} \propto \nu^{-2}$ for the later generations (we drop the subscripts labeling modes here for simplicity).  That is because at small detuning the term in brackets in Equations~(\ref{eq:Eth}) and (\ref{eq:Eth_harm}) is close to one and $E_{\rm th} \propto (\gamma/\kappa)^2\nu^{-2}$ with both $\gamma\propto \nu^{-2}$ and $\kappa \propto \nu^{-2}$ (Section~\ref{sec:method}). 

For the later  generations, there appear two tracks of well separated, albeit not substantially different, values of $E_{\rm th}$.  These correspond to the low and high $\ell$ modes present in each progeny pair when the $\ell$ of the progenitor mode is odd (since the sum of the three $\ell$ must be even in order to satisfy the angular selection rules of three-mode coupling).  The higher $\ell$ mode's minimum energy threshold tends to be set by the stochastic criterion (Equation~\ref{eq:Eth}) while the lower $\ell$ mode's threshold tends to be set by the harmonic criterion (Equation~\ref{eq:Eth_harm}).  However, as noted above, the overall results of our network integrations are not sensitive to this relatively small difference in threshold energies.

\section{\bf R\lowercase{esults}}
\label{sec:results}

As described in Section~\ref{sec:networks} and detailed in Table~\ref{tab:parameters}, we construct 24 mode networks in total, at each combination of $M=\{1.2, 2.0, 2.5\}M_\odot$, $\nu_{\rm max}\simeq\{50,100, 150,200\}\!\muHz$ and  $\gamma_p=\{10^{-8}, 10^{-7}\}\trm{ s}^{-1}$, where here $\gamma_p$ refers to the assumed damping rate of the $p$-$m$ parent modes.  We now present the results of integrating these networks.  We focus on how the nonlinear interactions impact the energetics of the parent modes because, as we describe in Section~\ref{sec:observations}, we can use the mode energy to calculate the observed mode visibility.  In Section~\ref{sec:prelim}, we first provide some additional information about our integration scheme.  In Section~\ref{sec:example_integration}, we describe the results of integrating the  $M2.0\nu98.4\gamma7.0$ network. The integration results of this network, whose structure we described in detail in Section~\ref{sec:example_network}, are representative of that of the other networks, whose results we show in Section~\ref{sec:suppression_all_networks}.

\begin{figure}[t!]
\centering
\includegraphics[width=3.4in]{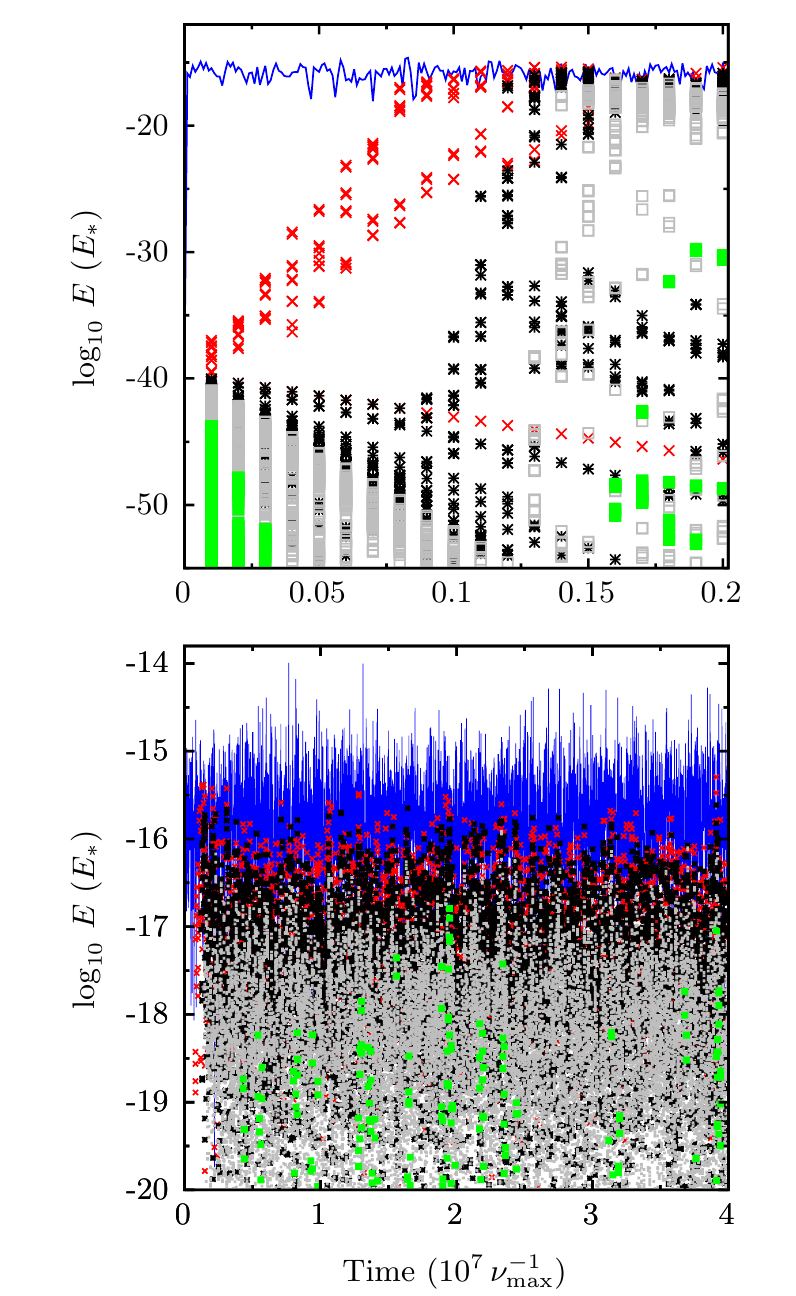} 
\caption{Energy as a function of time for a subset of the modes from the $M2.0\nu98.4\gamma7.0$ network. The top panel shows the dynamics at early times and the bottom panel  zooms in on energies above $10^{-20} E_\ast$. The modes are a $p$-$m$ parent mode (blue curve) and the cascade of modes it excites: daughters (red points), granddaughters (black points), great-granddaughters (gray points), and great-great granddaughters (green points). The parent is shown as a continuous blue curve  with a time resolution of  $10^4\numax^{-1} \simeq 3\trm{ yr}$, and the modes it excites are shown as discrete points with a time resolution of $10^5\numax^{-1}$. \label{fig:E_vs_t_all}}
\end{figure}

\begin{figure*}[t!]
\centering
\hspace{-1.5cm}
\includegraphics[width=7.0in]{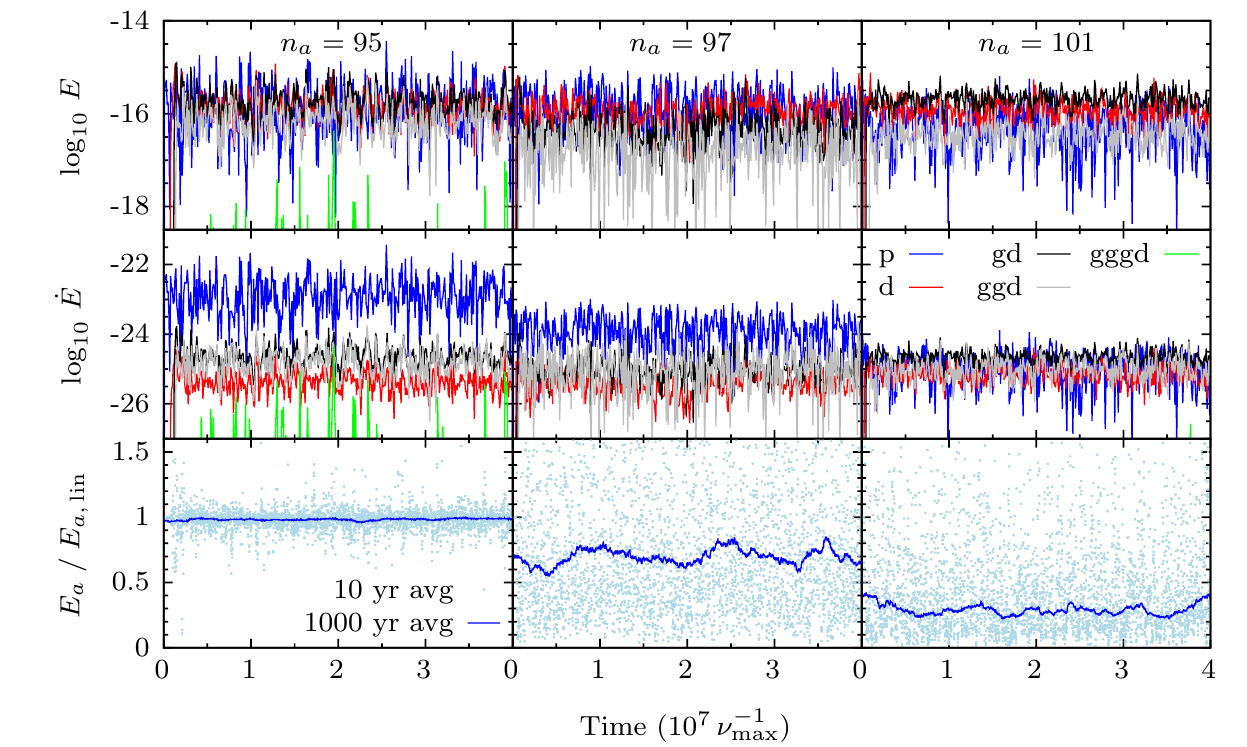} 
\caption{Evolution of each generations total mode energy $E$ (top row; units of $E_\ast$),  energy dissipation rate $\dot{E}$ (middle row; units of $E_\ast\trm{ s}^{-1}$), and the ratio of parent energy $E_a$ to its linear energy $E_{a, \rm lin}$ (bottom row), for a subset of the modes from the $M2.0\nu98.4\gamma7.0$ network.  The left, middle, and right columns show, respectively, the $n_a=95, 97, 101$ parent modes (blue lines; $\ell_a=m_a=1$ for all three). The $n_a=95$ parent is a $p$-$m$ mode with $\gamma_a\simeq10^{-7}\trm{ s}^{-1}$, while the other two parents are $g$-$m$ modes with $\gamma_a\simeq 10^{-8}\trm{ s}^{-1}$ and $2\times10^{-9}\trm{ s}^{-1}$, respectively. The top and middle rows also show the summed energy of the respective parent's daughters (red lines), granddaughters (black lines), great-granddaughters (gray lines), and great-great granddaughters (green lines). The  bottom row shows the moving average of $E_a/E_{a, \rm lin}$ over a duration of $10\trm{ yr}$ (light blue dots) and $1000\trm{ yr}$ (blue line).
\label{fig:Edot_sum}}
\end{figure*}

\subsection{Additional integration preliminaries}
\label{sec:prelim}

The modes in the networks are all given a negligibly small initial energy of $10^{-40} E_\ast$.  The linearly driven parents then very quickly reach their linear energies of $E_{a, \rm lin} \sim10^{-16} E_\ast$ (see Table~\ref{tab:parameters} for the exact values of $E_{a, \rm lin}$) and parametrically excite the daughters.  The daughters reach energies of $\sim10^{-17} E_\ast$ and parametrically excite the granddaughters which then excite the great-granddaughters and so on. We will see that for a typical network, a large fraction of the modes are eventually excited to energies $\gtrsim 10^{-18} E_\ast$.

The modes usually reach a nonlinear equilibrium by  $t=10^7\numax^{-1} \simeq 3000 (\numax / 100\muHz)^{-1} \trm{ yr}$ (we integrate each network for $4\times10^7\numax^{-1}$). At nonlinear equilibrium, some of the modes that were initially parametrically unstable fall below the threshold and their progeny decay away due to linear damping.   Since these progeny will, in reality, experience some degree of linear stochastic driving, we  artificially set their linear damping to zero whenever they decay below an energy of $10^{-60} E_\ast$; we find that the results are not sensitive to this essentially arbitrary choice of energy floor.

\subsection{Example network integration}
\label{sec:example_integration}

In Figure~\ref{fig:E_vs_t_all}, we show the energy as a function of time for a subset of the modes from the $M2.0\nu98.4\gamma7.0$ network. The modes are the $p$-$m$ parent mode $(\ell_a, m_a, n_a)=(1,1,95)$, and the modes it excites; to avoid making the plot too busy, for now we do not show the other parents of this network and the modes they independently excite.  The top panel shows the dynamics at early times. We see that the parent quickly drives the daughters to significant energies and the later generations are excited soon thereafter.  Within $t \approx 0.2 \times10^7 \numax^{-1}$, most of the later generation modes that will be excited have reached significant energies, while the unexcited modes decay away. 

The bottom panel of Figure~\ref{fig:E_vs_t_all} zooms in on higher energies  and shows the excited modes out to late times. At nonlinear equilibrium, the parent has the highest average energy while that of the other generations decreases in succession.  The great-great granddaughters (green points) are mostly unexcited, with the exception of brief intervals during which a few are momentarily driven to significant energies.  This is because at equilibrium the  great-granddaughters (gray points) have  $E\approx 10^{-18} E_\ast$, which is close to their parametric instability threshold (see Figure~\ref{fig:Eth_all_modes}).  We find that the great-granddaughters (but not the granddaughters) tend to be stable in our other networks as well, which explains why the results converge as long as we include at least four generations.

In Figure~\ref{fig:Edot_sum}, we show additional results from integrating the $M2.0\nu98.4\gamma7.0$ network. First  consider the left-column, which shows the same $p$-$m$ parent mode ($n_a=95$) as in Figure~\ref{fig:E_vs_t_all}, and the modes it excites.  Now, however, we show the total energy $E=\sum_a E_a$ and energy dissipation rate $\dot{E}=\sum_a \gamma_a E_a$, where the sums run over the modes in each generation.\footnote{Three-mode coupling contributes an additional $|\sum_{abc} \kappa_{abc} q_a q_b q_c|$ and $-2\sum_{abc} \gamma_a \kappa_{abc}\trm{Re}[q_aq_bq_c]$ to the total energy and dissipation rate, respectively.  However since $|\kappa_{abc}|$ is much smaller than the mode amplitudes, these contributions are negligible.} We see that the parent and the unstable generations all have nearly the same average total energy of $E\approx 10^{-16} E_\ast$, i.e., they are in equipartition with each other at a value near $E_{a, \rm lin}$. Their $\dot{E}$, on the other hand, are far from equal; for the $n_a=95$ parent, $\dot{E}\approx 10^{-23} E_\ast\trm{ s}^{-1}$ whereas for the other generations it excites $\dot{E}$ is one to two orders of magnitude smaller. That is because this parent is a $p$-$m$ mode for which convective damping is significant.  As a result, it has a large linear damping rate of $\gamma_a\simeq 10^{-7}\trm{ s}^{-1}$ and $\dot{E}_a=\gamma_a E_a \approx 10^{-23} E_\ast\trm{ s}^{-1}$.  By contrast, the other generations are  nonlinearly excited mode pairs that consist mostly of $g$-$m$ modes since they are more likely than $p$-$m$ modes to be parametrically unstable (due to the $g$-$m$ modes' smaller linear damping rates and  detunings).  For such modes, radiative-damping dominates and $\gamma_a\approx 10^{-9} - 10^{-8}\trm{ s}^{-1}$  (see Section~\ref{sec:linear_damping}); thus, since they are nearly in equipartition with the parent, their $\dot{E}$ are one to two orders of magnitude smaller.

The middle and right columns of Figure~\ref{fig:Edot_sum} show similar results but for $g$-$m$ parent modes ($n_a=97, 101$).   Just like with the $p$-$m$ parent mode shown in the left column, the generations are in energy equipartition with each other at $E\approx E_{a,\rm lin} \simeq 10^{-16} E_\ast$.  However, the $\dot{E}$ of these parent modes are, respectively, about ten times and fifty times smaller than the $\dot{E}$ of the $p$-$m$ parent mode, consistent with their smaller linear damping rates of $\gamma_a\simeq 10^{-8} \trm{ s}^{-1}$ and $2\times10^{-9}\trm{ s}^{-1}$ (see Figure~\ref{fig:gamma_parents}).   The $\dot{E}$ of the generations they excite, by contrast, are similar to those excited by the $p$-$m$ parent mode since they again consist of $g$-$m$ mode pairs with similar $E$ and linear damping rates.  We now describe how these features impact the energies of the individual parent modes.

\subsubsection{Suppression of the dipole-mode parents}
\label{sec:suppression_example_network}

\begin{figure}
\centering
\includegraphics[width=3.4in]{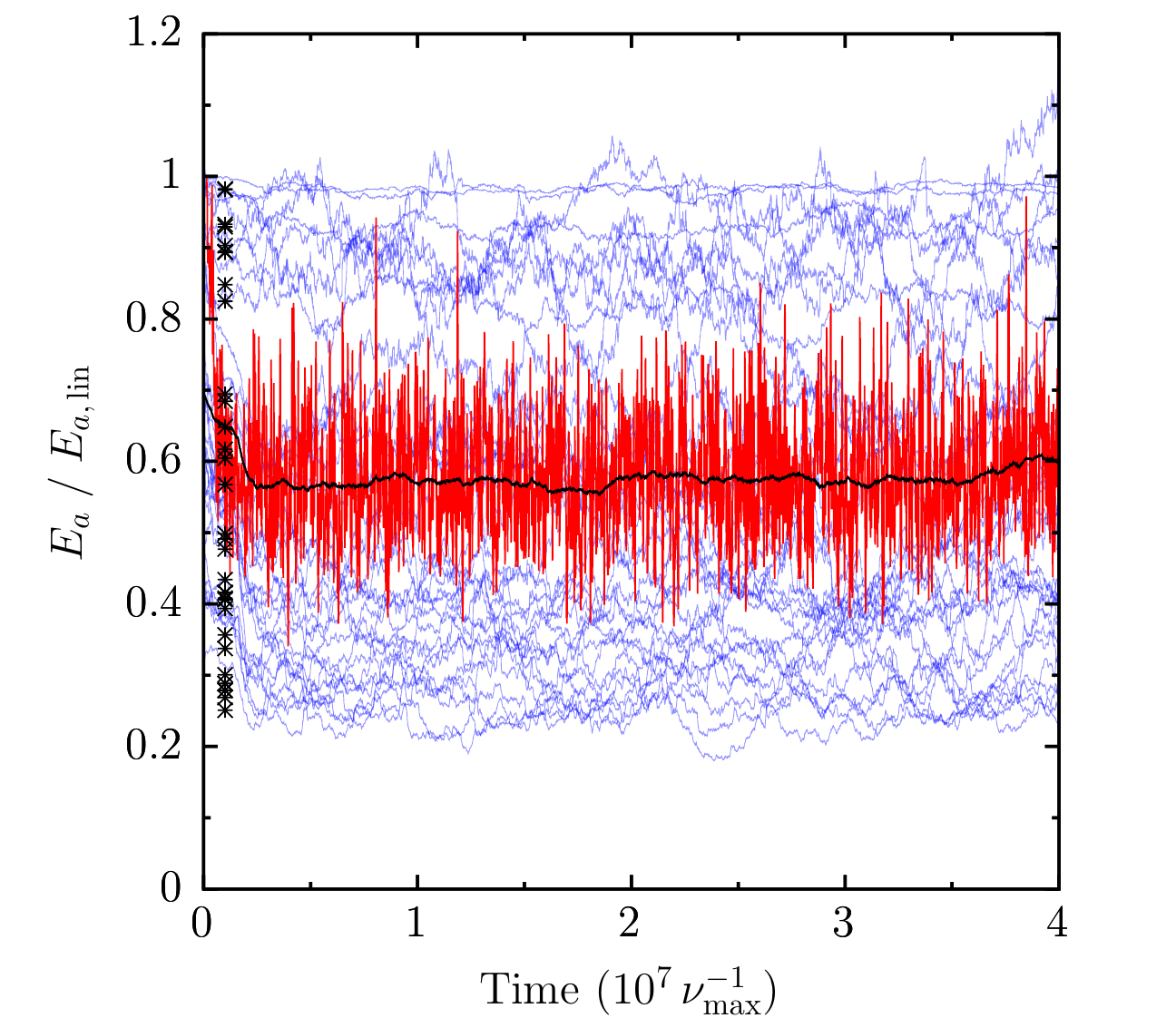} 
\caption{Ratio of parent mode energy $E_a$ to linear energy $E_{a, \rm lin}$ as a function of time for the $M2.0\nu98.4\gamma7.0$ network.  The blue lines show $E_a/E_{a, \rm lin}$ for each of the $N=33$ parent modes, averaged on thousand-year intervals.  The black points show the individual parents averaged over the full integration (arbitrary placed near $t=0$).  The red and black lines are the average over all the parents $\frac{1}{N}\sum E_a/E_{a, \rm lin}$, averaged on ten-year and thousand-year intervals, respectively. \label{fig:E_over_Elin_M2.0_nu100_gam7.0}}
\end{figure}

The bottom row of Figure~\ref{fig:Edot_sum} shows the ratio of the parent energy $E_a$ to its linear energy $E_{a, \rm lin}$, i.e., the energy it would have in the absence of nonlinear coupling.  To illustrate the magnitude of the fluctuations in $E_a/E_{a, \rm lin}$, we show results as a moving average over baselines of ten years and a thousand years.  Averaged over a thousand years, we find that the energy suppression factor $E_a/E_{a, \rm lin}$ is relatively constant, with $E_a/E_{a, \rm lin} \simeq 1, 0.7, 0.3$ for the $p$-$m$ parent mode and the two $g$-$m$ parent modes, respectively. However, averaged over ten years, which is representative of the baseline of observations (e.g., by \emph{Kepler}), $E_a/E_{a, \rm lin}$ fluctuates considerably.

For all of our networks, we find that the nonlinear interactions hardly suppress the amplitude of $p$-$m$ parent modes but substantially suppress the amplitudes of $g$-$m$ parent modes. To understand why, we can use an argument similar to one  \citet{Fuller:15} used to estimate amplitude suppression in red giants by the magnetic greenhouse effect. At nonlinear equilibrium,  there is a balance between the rate of energy input by stochastic driving $\dot{E}_{\rm in}$ and the rate of energy lost to thermal dissipation $\dot{E}_{\rm out}$.  The latter can be written as
\bea
\dot{E}_{\rm out} &=& \gamma_a E_{a, \rm eq} + \sum_{b\neq a} \gamma_b E_{b, \rm eq}
\non &=& \left(\gamma_a +\Gamma_{a, \rm nl}\right) E_{a, \rm eq}, 
\eea
where $E_{a, \rm eq}$ is the equilibrium energy of the parent and $\Gamma_{a, \rm nl} \equiv \sum_{b\neq a} \gamma_b E_{b, \rm eq} / E_{a, \rm eq}$ is the effective nonlinear damping rate of the parent at equilibrium due to all the modes it nonlinearly excites.    If we assume that the power supplied to the parent by stochastic driving is unaltered by the nonlinear interactions,\footnote{\citet{Fuller:15} make this assumption as well, although it is not clear that it necessarily holds.  On the one hand, in both cases the suppression mechanism (nonlinear interactions or the magnetic greenhouse effect) is localized to the core whereas the stochastic driving is localized to the convective envelope.  On the other hand, the power supplied to a mode by stochastic driving may have a nonlinear dependence on the mode's energy.} 
then $\dot{E}_{\rm in}=\gamma_a E_{a, \rm lin}$.  Since $\dot{E}_{\rm out}=\dot{E}_{\rm in}$ at equilibrium, we have
\beq
\frac{E_{a, \rm eq}}{E_{a, \rm lin}} = \frac{\gamma_a}{\gamma_a + \Gamma_{a, \rm nl}}.
\label{eq:Ea_Ealin_approx}
\eeq
From the numerical results presented in Figure~\ref{fig:Edot_sum}, we see that $\Gamma_{a, \rm nl} \approx 10^{-25} E_\ast\trm{ s}^{-1} /10^{-16} E_\ast \approx 10^{-9}\trm{ s}^{-1}$ for all three parent modes shown.  Thus, for the $p$-$m$ parent mode $\gamma_a \gg \Gamma_{a, \rm nl}$ and $E_{a, \rm eq}/{E_{a, \rm lin}}\simeq 1$, while for the $g$-$m$ parent modes $\gamma_a \approx \Gamma_{a, \rm nl}$ and $E_{a, \rm eq}/{E_{a, \rm lin}} \lesssim 1$.  

Why is $\Gamma_{a, \rm nl} \approx 10^{-9}\trm{ s}^{-1}$ for all three parents?  We expect the nonlinear damping rate of the parent  to depend on how effectively it excites daughter modes.  When the parent is well above the stochastic parametric threshold energy, the initial growth rate of the fastest growing daughter mode pair is (\citetalias{Weinberg:19})
\bea
\Gamma_{a, \rm param} &\simeq& \frac{\omega_b \omega_c \kappa_{abc}^2}{\gamma_a}   \frac{E_a}{ E_\ast}
\non &\approx& 1.0\times 10^{-10} \left(\frac{\nu_a}{100\muHz}\right)^{-2} 
\non && \times \left(\frac{\gamma_a}{10^{-7}\trm{ s}^{-1}}\right)^{-1} \left(\frac{E_a}{10^{-16}E_\ast}\right) \trm{ s}^{-1},
\eea
where in the numerical expressions we assume $|\omega_b| \simeq |\omega_c| \simeq  |\omega_a/2|$ and we used the fact that $\kappa_{abc}\simeq 10^3 (\nu_a/100\muHz)^{-2}$ (Section~\ref{sec:method}). Although $\Gamma_{a, \rm nl} \approx 10 \times \Gamma_{a, \rm param}$,  \citet{Essick:16} and \citet{Yu:20} also find that the nonlinear dissipation rate is $\approx 10$ times larger than the parametric growth rate in their nonlinear mode network calculations of the dynamical tide in hot Jupiter and white dwarf binary systems, respectively. This is likely because there are many daughter modes excited by a parent rather than just a single pair. It therefore seems reasonable that we find $\Gamma_{a, \rm nl}\approx 10^{-9}\trm{ s}^{-1}$ for all three parents.

Figure~\ref{fig:E_over_Elin_M2.0_nu100_gam7.0} shows the energy suppression factor $E_a/E_{a, \rm lin}$ of each of the thirty-three parents of the $M2.0\nu98.4\gamma7.0$ network.  The blue lines are thousand-year averages and the black points near the $t=0$ axis are the average over the full integration.  The  suppression of individual parents range from $E_a/E_{a, \rm lin}\simeq 0.2$ to 1.0 going from the $g$-$m$ modes with the smallest $\gamma_a\simeq 10^{-9}\trm{ s}^{-1}$ to the $p$-$m$ modes with the largest $\gamma_a\simeq 10^{-7}\trm{ s}^{-1}$ (corresponding to the minima and maxima of the blue $\gamma_a$ curve in  Figure~\ref{fig:gamma_parents}).  

Figure~\ref{fig:E_over_Elin_M2.0_nu100_gam7.0} also shows $\frac{1}{N}\sum E_a/E_{a, \rm lin}$, the suppression  averaged over the $N=33$ parents. It is about $0.6$ on a thousand year interval (black line), with $\approx 20\%$ fluctuations on ten year intervals (red line).

\subsection{Suppression as a function of  $\numax$ and $M$}
\label{sec:suppression_all_networks}

\begin{figure*}[t!]
\centering
\includegraphics[width=7.3in]{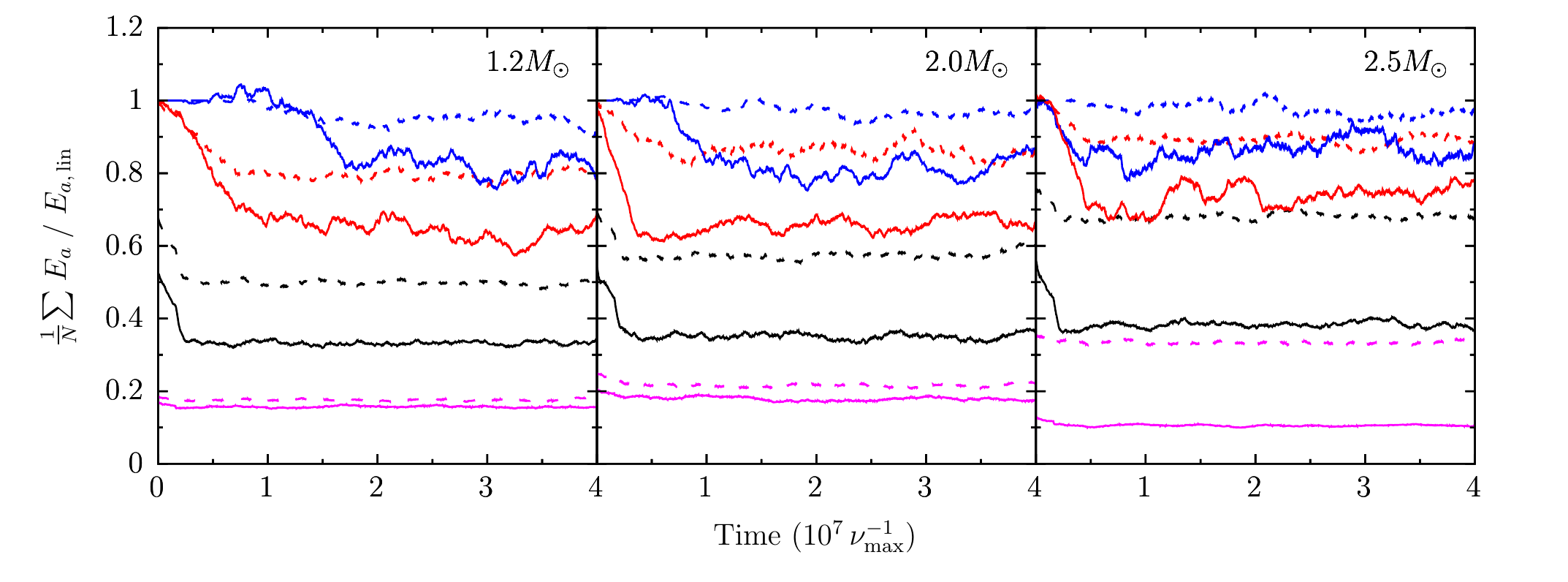} 
\caption{Ratio of parent energy to linear energy averaged over all the parents in a network $\frac{1}{N}\sum E_a/E_{a, \rm lin}$ as a function of time for each of the 24 networks.  The averages are taken on thousand-year intervals.  The left, middle, and right panels show results for the $M=\{1.2, 2.0, 2.5\} M_\odot$ networks, respectively.  In each panel, the results from top to bottom are the $\numax\simeq \{200, 150, 100, 50\}\mu\trm{Hz}$ networks, shown as blue, red, black, and cyan lines, respectively. Solid lines (dashed lines) are networks whose $p$-$m$ mode has a damping rate $\gamma_p\simeq 10^{-8}\trm{ s}^{-1}$ ($\gamma_p \simeq 10^{-7}\trm{ s}^{-1}$). \label{fig:E_over_Elin_all_networks}}
\end{figure*}

In Figure~\ref{fig:E_over_Elin_all_networks} we show the average energy suppression of the parent modes $\frac{1}{N}\sum E_a/E_{a, \rm lin}$ as a function of time for each of our 24 networks.  We find that the degree of suppression increases significantly with decreasing  $\numax$.  Specifically, at $\numax\simeq \{200, 150, 100, 50\}\muHz$ the parents are suppressed by $\{10-20, 20-40, 40-70, 80-90\}\%$, respectively. 

The main reason the suppression increases significantly with decreasing $\numax$ is that at smaller $\numax$ there are many more $g$-$m$ parent modes between each $p$-$m$ parent mode.  We saw in Section~\ref{sec:suppression_example_network} that the $g$-$m$ parents are much more strongly suppressed owing to their smaller linear damping rates (Equation~\ref{eq:Ea_Ealin_approx}). By contrast, the $p$-$m$ modes have a comparatively large $\gamma_a$ (due to strong convective damping) and are therefore nearly unsuppressed at all
$\numax$.  Thus, given the higher proportion of $g$-$m$ modes at small $\numax$, the value of the average suppression $\frac{1}{N}\sum E_a/E_{a, \rm lin}$ is much smaller.

A secondary, but much less significant, reason for the $\numax$ dependence is that at smaller $\numax$ the parents' linear driving pushes them higher above the parametric threshold $E_{a, \rm th}$ (e.g., compare the top and bottom panels of Fig.~\ref{fig:Eth_parent}). Thus, the parents drive the daughter modes more vigorously at smaller $\numax$. This does not, however, significantly impact  the nonlinear equilibrium nor the degree of parent mode suppression.  To see an illustration of this, notice that at $\numax\simeq 200\muHz$,  $E_{a,\rm lin}$ is much larger for the $M=2.5M_\odot$ model than for the $M=1.2 M_\odot$ and $2.0 M_\odot$ models (due its higher $T_{\rm eff}$; see Table~\ref{tab:parameters}), and yet the suppression is comparable across the three masses (blue lines in Figure~\ref{fig:E_over_Elin_all_networks}).  

We find that the suppression is fairly insensitive to $M$.  This is because $\kappa_{abc}$, the linear damping rates, and frequency detunings are not sensitive to $M$ (Figure~\ref{fig:gamma_kappa}).  The suppression also does not depend strongly on the assumed convective damping rate  $\gamma_p\simeq \gamma_{a,\rm conv}$  of the $p$-$m$ modes (compare the solid and dashed lines in Figure~\ref{fig:E_over_Elin_all_networks}). Although the $p$-$m$ modes are more suppressed when  $\gamma_p=10^{-8}\trm{ s}^{-1}$ rather than $10^{-7}\trm{ s}^{-1}$, only a small subset of the modes in the sum $\frac{1}{N}\sum E_a/E_{a, \rm lin}$ are $p$-$m$ modes.  Most  are $g$-$m$ modes and they are primarily damped by radiative damping and thus  insensitive to the assumed contribution of convective damping (since their mode inertia $\mathcal{M}_a$ are much larger; see Section~\ref{sec:linear_damping}).

\section{\bf C\lowercase{omparison to} O\lowercase{bservations}}
\label{sec:observations}

\subsection{Visibility of dipole modes}

In order to compare our results with observations, we need to relate our calculations of parent mode energies to dipole mode visibilities 
\beq
V^2(\ell=1) = \mathlarger{\sum_{\ell_a=1}} \frac{\langle A_a^2 \rangle}{\langle A_0^2\rangle},
\eeq
where $\langle A_a^2\rangle$ and $\langle A_0^2 \rangle$ are the observed mean square amplitudes of a dipole mode and its neighboring radial mode and the sum runs over all dipole modes between consecutive radial mode peaks  (\citealt{Mosser:12:depressed, Stello:16a}; in practice, these papers integrate the observed power over a $4\Delta\nu$-wide range of the spectrum centred on $\numax$, where $\Delta \nu$ is the large frequency separation).  The amplitude of a mode, like the mean square surface velocity $\langle  v_a^2\rangle$, is proportional to the integral of its power spectrum and thus $\langle A_a^2 \rangle \propto \langle  v_a^2\rangle = E_a / \mathcal{M}_a$. The dipole mode visibility is therefore given by
\beq
V^2 = \mathlarger{\sum_{\ell_a=1}} 
\frac{E_a\mathcal{M}_0}{E_0\mathcal{M}_a}.
\label{eq:visibility_general}
\eeq
In the limit that all the modes are in energy equipartition ($E_a=E_0$),
\beq
V^2 = \mathlarger{\sum_{\ell_a=1}}  \frac{\mathcal{M}_0}{\mathcal{M}_a} \simeq 1
\label{eq:normal_visibility}
\eeq
(see \citet{Mosser:17} for the reason the sum of the radial mode to dipole mode inertia  approximately equals one).

These expressions do not account for bolometric and geometric corrections that arise when comparing dipole and radial mode amplitudes. \citet{Ballot:11} show (see also \citealt{Mosser:11, Mosser:17}) that these modify the dipole mode visibility by a nearly constant factor of $C\simeq 1.54$ for red giants observed by \emph{Kepler}.

The dipole mode visibility of \emph{Kepler} red giants show two populations: those with normal visibility and those with suppressed visibility. The former is considered normal because their $V^2\simeq 1.5$ \citep{Mosser:12:depressed, Mosser:17, Stello:16a}, consistent with Equation (\ref{eq:normal_visibility}) after accounting for the correction factor $C$.  This suggests that in red giants with normal visibility, the dipole and radial modes are in energy equipartition.  In this case, on average $E_a=E_0$, although at any given moment the energies will not be exactly equal due to the stochastic nature of the driving.  We therefore define the normal dipole mode visibility as
\beq
V^2_{\rm norm}= C\mathlarger{\sum_{\ell_a=1}}  \frac{E_{a, \rm lin}}{E_0}\frac{\mathcal{M}_0}{ \mathcal{M}_a} \simeq 1.5.
\eeq
Here we set $E_a=E_{a,\rm lin}$ in Equation~(\ref{eq:visibility_general}) because $E_a\simeq E_0$ (normal dipole modes are in equipartition with radial modes) and $E_0\simeq E_{a, \rm lin}$ (radial modes are unaffected by nonlinear damping).\footnote{We do not define the normal visibility as $V^2_{\rm norm}= C \sum_{\ell_a=1}  \mathcal{M}_0 /\mathcal{M}_a$ because although in equipartition $E_0\simeq E_{a, \rm lin}$ on average, we are interested in not just the mean value of $V^2_{\rm norm}$ (which we know is about 1.5), but also the dispersion given the finite duration of observations.}   Radial modes should be unaffected by nonlinear damping  because their displacements $\gv{\xi}_0$ near the stellar center are much smaller  than that of dipole modes.  Their nonlinear coupling to other modes is therefore weaker and their $\kappa_{abc}$ values are smaller.  Moreover, radial modes have comparatively large damping rates  due to strong convective damping.  Thus, just   like we found for the dipole $p$-$m$ modes, convective damping of the radial modes should overwhelm their nonlinear damping, such as it is (i.e., their $\gamma_a \gg \Gamma_{a, \rm nl}$ in Eq.~\ref{eq:Ea_Ealin_approx}).

By contrast, the red giants with suppressed visibility have $V^2 \ll V^2_{\rm norm}\simeq 1.5$. From Equation~(\ref{eq:visibility_general}) it follows that this is because the mode energies are not all in equipartition ($E_a < E_0$).  We therefore define the suppressed dipole mode visibility as
\beq
V^2_{\rm sup}=
C\mathlarger{\sum_{\ell_a=1}} 
\frac{E_a \mathcal{M}_0}{E_0 \mathcal{M}_a}.
\eeq
The ratio of visibility between suppressed and normal dipole modes, which we will refer to as the normalized visibility $\bar{V}^2$, is therefore
\beq
\bar{V}^2 \equiv \frac{V^2_{\rm sup}}{V^2_{\rm norm}} 
= \dfrac{\mathlarger{\sum_{\ell_a=1}}
E_a \mathcal{M}_a^{-1}}
{\mathlarger{\sum_{\ell_a=1}} E_{a, \rm lin} \mathcal{M}_a^{-1}}.
\label{eq:Vsup_to_Vnorm}
\eeq
Note that this expression does not depend on the properties of the radial modes (having assumed that they are unaffected by nonlinear damping). 

\subsection{Comparing the computed and observed dipole mode visibilities}
\label{sec:comparing_visibilities}

\begin{figure*}[t!]
\hspace{-0.0cm}
\includegraphics[width=7.0in]{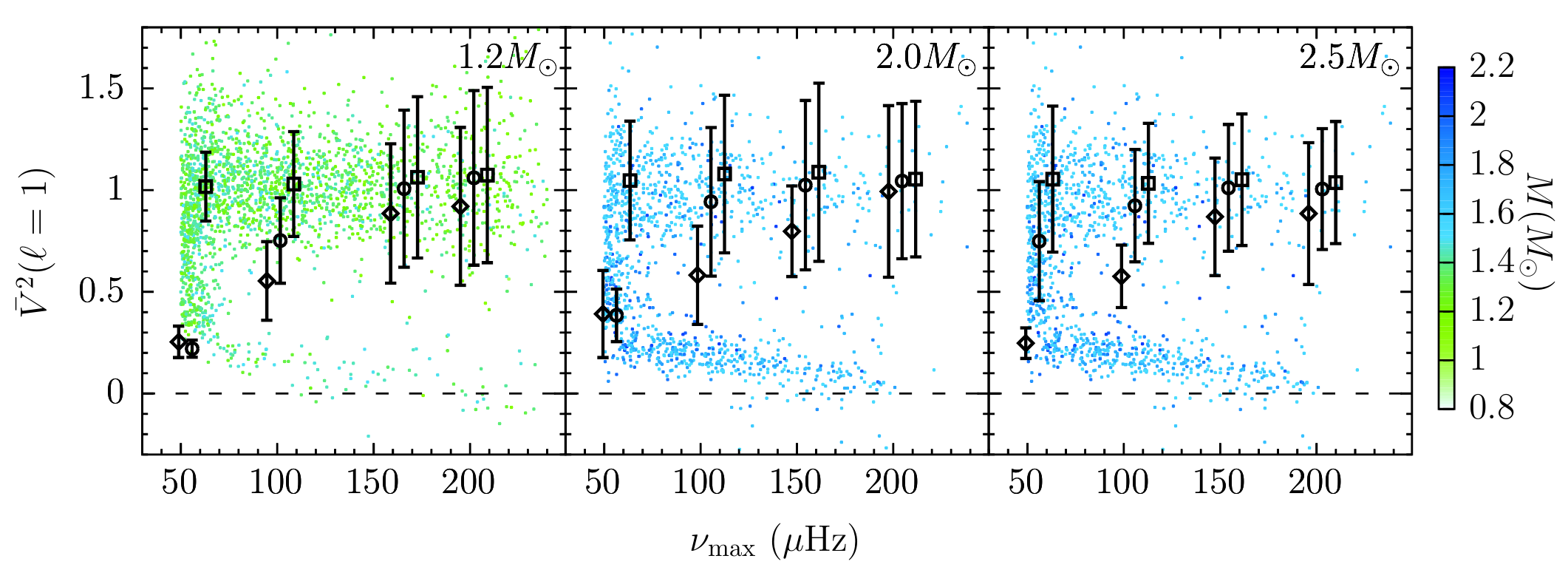} 
\caption{Normalized dipole mode visibility $\bar{V}^2$ (Equation~\ref{eq:Vsup_to_Vnorm}) as a function $\numax$ from the network integrations (black points) at $M=\{1.2, 2.0, 2.5\} M_\odot$ (right, middle, and left panels).  The black diamonds (circles) show the networks whose parent $p$-$m$ mode has a damping rate  $\gamma_p\simeq 10^{-8}\trm{ s}^{-1}$ ($10^{-7}\trm{ s}^{-1}$), while the black squares show  $\bar{V}^2$ in the absence of nonlinear damping.   The error bars are the standard deviation of the ten-year moving average. For clarity, the black circles and squares are shifted slightly from their model's $\numax$.   The colored points are the observed dipole-mode visibilities of \emph{Kepler} red giants taken from \citet{Stello:16b}, divided by 1.35, the mean visibility of the normal stars \citep{Stello:16a}.  \emph{Kepler} red giants with $M<1.5 M_\odot$ ($M\ge 1.5 M_\odot$) are shown in the left panel (middle and right panels; see color scale on the right).
\label{fig:visibility}}
\end{figure*}

In order to compute the normalized dipole mode visibility $\bar{V}^2$, we  take the energy $E_a(t)$ and linear energy $E_{a, \rm lin}(t)$ of the parent modes from our network integrations and the  mode masses $\mathcal{M}_a$ from \texttt{GYRE} and calculate the moving averages of the sums in the numerator and denominator\footnote{\label{footnote:Ealin_t_shift} In our integrations, the total energy of a mode $E_a(t)$ can be strongly correlated in time with its linear energy $E_{a, \rm lin}(t)$, especially for the $p$-$m$ parent modes that are only slightly affected by nonlinear damping. Therefore, in order to obtain a realistic estimate of the \emph{dispersion} of $\bar{V}^2$, we insert an arbitrary time shift $E_{a,\rm lin}(t) \rightarrow E_{a,\rm lin}(t+t_0)$ when evaluating the denominator of Equation~(\ref{eq:Vsup_to_Vnorm}). This does not affect the mean value of $\bar{V}^2$ and the dispersion is insensitive to $t_0$ as long as it is not  too small.} of Equation~(\ref{eq:Vsup_to_Vnorm}). We show the results in Figure~\ref{fig:visibility} (black points) and compare them with the observed visibility of dipole modes in \emph{Kepler} red giants (colored points; \citealt{Mosser:12:depressed, Mosser:17, Stello:16a, Stello:16b}). 

\subsubsection{$\numax \ge 100\muHz$}

First consider the results of the  $\numax \ge 100\muHz$ models. We find that networks whose parent $p$-$m$ mode has a linear damping rate $\gamma_p\simeq 10^{-7}\trm{ s}^{-1}$ (black circles) have an average normalized visibility  $\bar{V}^2\simeq 1$ for $\numax = 150-200\muHz$ and  $\bar{V}^2\simeq 0.8-1$ for $\numax \simeq 100\muHz$.  Networks with $\gamma_p\simeq 10^{-8}\trm{ s}^{-1}$ (black diamonds) have $\bar{V}^2\simeq 0.9-1$ for $\numax = 150-200\muHz$  and $\bar{V}^2\simeq 0.6$ for $\numax \simeq 100\muHz$.  As with the energy ratio results (Figure~\ref{fig:E_over_Elin_all_networks}), there is only a weak dependence on $M$. For reference, we also show $\bar{V}^2$ in the absence of nonlinear damping, obtained by setting $E_a(t)=E_{a,\rm lin}(t)$ in the numerator of Equation~(\ref{eq:Vsup_to_Vnorm}; see footnote~\ref{footnote:Ealin_t_shift}). In that case,  $\bar{V}^2\simeq 1$ at all $\numax$, as expected.   The error bars, which correspond to the standard deviation of the ten-year moving average, extend $\approx 20-30\%$ about the average on all three types of points. The observed spread in $\bar{V}^2$  could therefore be explained by the stochastic nature of the fluctuations averaged over the four years of \emph{Kepler} data.

The \emph{Kepler} stars with suppressed dipole modes appear in Figure~\ref{fig:visibility} as a band of colored points with  $\bar{V}^2 \lesssim 0.2$  extending out to high $\numax$,  especially for $M\gtrsim 1.3 M_\odot$.\footnote{Some of the stars analyzed by \citealt{Stello:16a}  have $\bar{V}^2<0$, which is unphysical. They note that this is because the uncertainty in their background-corrected spectra introduces measurement scatter on top of the intrinsic spread in the visibilities.}  Nonlinear damping is clearly unable to account for these stars.  At most, it may be able to account for the few stars with $\bar{V}^2\approx 0.5$ near $\numax \approx 100\muHz$.  However, this would require $\gamma_p\simeq 10^{-8}\trm{ s}^{-1}$, which, as we explain in Section~\ref{sec:linear_damping}, is on the low side of likely $\gamma_p$ values.  For the more typical value of $\gamma_p\simeq 10^{-7}\trm{ s}^{-1}$,  nonlinear damping has only a mild influence on $\bar{V}^2$ at high $\numax$.

The computed $\bar{V}^2$ at  $\numax \ge 100\muHz$ are noticeably larger, i.e., closer to one, than the average ratio of parent energy to linear energy  $\frac{1}{N}\sum E_a / E_{a, \rm lin}$ (Figure~\ref{fig:E_over_Elin_all_networks}).   This is because at high $\numax$, the sums in Equation~(\ref{eq:Vsup_to_Vnorm}) are dominated by the $p$-$m$ modes owing to their small inertias (large $\mathcal{M}_a^{-1}$). Since the $p$-$m$ modes all have $E_a \simeq E_{a, \rm lin}$ (Section~\ref{sec:suppression_example_network}), the $\mathcal{M}_a^{-1}$-weighted sums over energy in Equation~(\ref{eq:Vsup_to_Vnorm})  ensure that $\bar{V}^2 \simeq 1$ even when $\frac{1}{N}\sum E_a / E_{a, \rm lin} \lesssim 1$.\footnote{\label{footnote: inertias} More quantitatively, the inertia ratio of $g$-$m$ to $p$-$m$ mode neighbors is $\mathcal{M}_p/\mathcal{M}_g \simeq (1+N/4)/(1+4N)$ (see Appendix A of \citealt{Grosjean:14}).  Thus, the sum between acoustic peaks $\sum \mathcal{M}_g^{-1} \simeq N (1+N/4)/(1+4N)\mathcal{M}_p^{-1}$. For $\numax \gtrsim 100\muHz$, $N \lesssim 10$ and this sum is somewhat less than $\mathcal{M}_p^{-1}$ (i.e., the $p$-$m$ mode dominates) while for $\numax \lesssim 50\muHz$, $N \gtrsim 50$ and the sum equals a few $\mathcal{M}_p^{-1}$ (the $g$-$m$ modes dominate).}

\subsubsection{$\numax \simeq 50\muHz$}

For more evolved red giants with $\numax \simeq 50\muHz$, we find that nonlinear damping has a dramatic influence on $\bar{V}^2$, unlike at high $\numax$.  For both values of $\gamma_p$, we find $\bar{V}^2 \simeq 0.2-0.4$ (the one exception is at $\gamma_p=10^{-7}\trm{ s}^{-1}$ and $M=2.5 M_\odot$ for which  $\bar{V}^2\simeq 0.7$).  We also see that $\bar{V}^2$ approximately equals the values  of $\frac{1}{N}\sum E_a / E_{a, \rm lin}$ shown in Figure~\ref{fig:E_over_Elin_all_networks}; this is because at small $\numax$, the sums in Equation~(\ref{eq:Vsup_to_Vnorm}), like $\frac{1}{N}\sum E_a / E_{a, \rm lin}$,  are dominated by the $g$-$m$ modes and not the $p$-$m$ mode (see footnote~\ref{footnote: inertias}).

The calculated $\bar{V}^2$ at $\numax\simeq 50\muHz$ are all small whereas the \emph{Kepler} stars span the full range $0 \lesssim \bar{V}^2\lesssim 1$.  This discrepancy could be explained if the observed stars mostly have $\gamma_p > 10^{-7}\trm{ s}^{-1}$ and thus experience weaker nonlinear damping than the networks we consider (see Section~\ref{sec:suppression_example_network} and Equation~\ref{eq:Ea_Ealin_approx}).  However, the estimates in Section~\ref{sec:linear_damping} suggest that the stars are unlikely to have a $\gamma_p$  much larger than $2\times 10^{-7}\trm{ s}^{-1}$.  Another possibility is that because of strong radiative damping, the later generations are traveling waves rather than the standing waves we assume in our analysis.  In particular, for $\nu_a \lesssim 30\muHz$, a dipole $g$-$m$ mode has such a short wavelength in the core that  radiative damping  can  dissipate all of its energy  in less than its group travel time across the star \citep{Dupret:09, Grosjean:14}. Such modes are therefore in going traveling waves rather than standing waves.  Since the parent has a frequency near $\numax \simeq 50\muHz$, the granddaughters and great-granddaughters all have frequencies well below $30\muHz$ and should therefore be traveling waves not  standing waves. If the nonlinear mode dynamics  involve a standing wave exciting traveling waves, the true nonlinear equilibrium might be different from the one we found, and could lead to a large spread in $\bar{V}^2$.

\subsection{Influence of nonlinear damping on  linewidths}

In Section~\ref{sec:suppression_example_network}, we found that the effective nonlinear damping rate of the parent modes is $\Gamma_{a, \rm nl} \sim 10^{-9}\trm{ s}^{-1}$ (even at high $\numax$).  While such damping is well below the linear damping rate of the $p$-$m$ modes, it can exceed the linear damping rate of the $g$-$m$ modes by a factor of $\sim 10$ and thus be their dominant source of dissipation. Nonlinear damping should therefore broaden the peaks of $g$-$m$ modes in the observed power spectra.  Such damping corresponds to a linewidth of $\pi \Gamma_{a, \rm nl} \sim 3\trm{ nHz}$, which is close to  \emph{Kepler}'s  resolution limit of  $2/\pi T_{\rm obs} \simeq 5\trm{ nHz}$ \citep{Dupret:09} given its baseline of $T_{\rm obs}\simeq 4\trm{ yr}$.

In their peak bagging analysis of 19 \emph{Kepler} stars, \citet{Corsaro:15} attempted to measure  the linewidths of $g$-$m$ dipole modes near $\numax$.  However, they found that the majority of these modes are unresolved. Of the resolved ones, the smallest has a measured linewidth of $\approx 30\trm{ nHz}$.  This suggests that \emph{Kepler}'s resolution limit of $5\trm{ nHz}$ is still somewhat too large to resolve the influence of nonlinear damping on the $g$-$m$ mode linewidths.

\section{\bf S\lowercase{ummary} \lowercase{and} C\lowercase{onclusions}}
\label{sec:summary}

We studied the impact of nonlinear mode coupling on mixed modes in red giants.  In \citetalias{Weinberg:19}, we showed that the stochastic forcing of dipole modes  by turbulent motions in the convective envelope drive the modes to such large energies that they are unstable to weakly nonlinear, resonant three-wave interactions.  Motivated by those results, here we analyzed the time-dependent evolution of  unstable dipole modes over a broad range of stellar mass and evolutionary state. This entailed solving the amplitude equations for large networks of nonlinearly coupled modes.  The networks consisted of primary (i.e., parent) dipole modes, which are directly excited by  stochastic forcing and span an acoustic peak (from $p$-$m$ to $g$-$m$ modes), coupled to several generations of secondary modes with low instability thresholds.  
We constructed 24 networks, consisting of several thousand modes each, across a  grid of models designed to span the range of red giants observed by \emph{Kepler} and account for uncertainties in the $p$-$m$ mode (convective) damping rate.

By integrating each network for more than $10^7$ parent mode periods (several thousand years), we determined the nonlinear equilibria of the parents and the secondary modes they excite. We found that the parents' equilibrium energies were suppressed relative to their linear values ($E_{a, \rm eq} < E_{a,\rm lin}$), with the degree of suppression  sensitive to whether the parent was a $p$-$m$  or $g$-$m$ mode; the former were hardly suppressed at all whereas the latter could be suppressed by factors of five to ten (Figure~\ref{fig:E_over_Elin_M2.0_nu100_gam7.0}). We argued that this is because $E_{a, \rm eq}/E_{a,\rm lin} \approx \gamma_a / (\gamma_a + \Gamma_{a, \rm nl})$, where $\gamma_a$ and $\Gamma_{a, \rm nl}$ are the linear and effective nonlinear damping rates of the parent modes, respectively.   Whereas $\Gamma_{a,\rm nl}$ is nearly the same for both types of modes,  $\gamma_a$ is  $\sim 10- 100$ times larger for $p$-$m$ modes due to their small inertias (Figure~\ref{fig:gamma_parents}). As a result, we found that for the $g$-$m$ modes $\Gamma_{a,\rm nl}\gtrsim \gamma_a$ and $E_{a, \rm eq}/E_{a,\rm lin} \lesssim 0.2$, while for the $p$-$m$ modes $\Gamma_{a,\rm nl} \ll \gamma_a$ and $E_{a, \rm eq}/E_{a,\rm lin} \simeq 1$.

Since nonlinear damping is the dominant source of dissipation of $g$-$m$ modes, it should determine their linewidths in the observed power spectra (even at higher $\numax$). We found $\Gamma_{a, \rm nl} \sim 10^{-9}\trm{ s}^{-1}$, which corresponds to a linewidth of $\sim 3\trm{ nHz}$. This is close to \emph{Kepler}'s resolution limit of $5\trm{ nHz}$, although the peak bagging analysis of 19 \emph{Kepler} red giants by \citet{Corsaro:15} suggests that a resolution of $5\trm{ nHz}$ is only sufficient to measure linewidths $\gtrsim 30\trm{ nHz}$. 

In order to further compare our results with  \emph{Kepler} asteroseismic measurements, we computed the normalized visibility of the parent dipole modes $\bar{V}^2$, which we showed equals the inertia-weighted sums over the parent's $E_{a,
\rm eq}$ and $E_{a, \rm lin}$  (Equation~\ref{eq:Vsup_to_Vnorm} and Figure~\ref{fig:visibility}).  We found that for $\numax \gtrsim 100\muHz$, nonlinear damping has only a mild effect on the visibility, yielding $\bar{V}^2\simeq 0.8-0.9$ (compared to the linear value  $\bar{V}^2\simeq 1$).   For more evolved red giants with $\numax \lesssim 100\muHz$, nonlinear damping's effect is much more significant, with $\bar{V}^2\simeq 0.2$ at $\numax=50\muHz$ (still with a weak $M$ dependence). The principal reason for this difference is that in more evolved red giants there are many more $g$-$m$ modes between acoustic peaks.  They therefore dominate the sum over modes in $\bar{V}^2$ and, because of their small $E_{a, \rm eq}/E_{a,\rm lin}$, weigh down its value. A caveat is that by $\numax\approx 50\muHz$, the later generation modes are likely traveling waves rather than standing waves (due to strong radiative damping), in which case our standing wave treatment of mode coupling is no longer appropriate.

Observations show that a significant fraction of \emph{Kepler} red giants with $M\gtrsim 1.3 M_\odot$ and high $\numax$ have dipole modes with $\bar{V}^2 \lesssim 0.2$ (\citealt{Mosser:12:depressed, Mosser:17, Stello:16a, Stello:16b}; see Figure~\ref{fig:visibility}). Given that we find only weak suppression at high $\numax$, we conclude that resonant mode interactions cannot account for these suppressed dipole mode stars. At smaller $\numax\approx 50\muHz$, the observed visibilities are  spread across the full range of values ($0 \lesssim \bar{V}^2 \lesssim 1$).  Our results indicate that nonlinear damping can have a significant impact on the dipole modes in these more evolved stars, and interpreting their $\bar{V}^2$ likely requires accounting for the excitation of nonlinear secondary waves in the traveling wave regime. 

Our analysis only considered \emph{resonant} three-wave interactions, i.e., parametric instabilities, in which a parent  ($a$) excites pairs of daughters $(b,c)$ that satisfy $\omega_b+\omega_c\simeq \omega_a$. A different type of three-wave interaction involves a turbulently driven parent coupling to itself (or, more generally, another parent) and driving a daughter at twice its frequency $\omega_b \simeq 2\omega_a$.  This is a form of (nonlinear) inhomogeneous driving, as the nonlinear term in the daughter amplitude equation depends on $\kappa_{aab} (q_a^\ast)^2 \propto e^{2i\omega_a t}$.  Importantly, if $\omega_b > \omega_{\rm ac}$, where $\omega_{\rm ac}$ is the acoustic cutoff frequency, then the daughter propagates above the photosphere and dissipates all if its energy there rather than forms a standing wave. Since the frequency of maximum power $\omega_{\rm max}=2\pi \nu_{\rm max} \simeq 0.6 \omega_{\rm ac}$ \citep{Belkacem:11}, if $\omega_a  \simeq \omega_{\rm max}$, then $\omega_b \ga \omega_{\rm ac}$.   Previous studies \citep{Kumar:89, Kumar:94} have pointed out that such an upward going traveling wave can give rise to a substantial loss of energy from the system of modes within the star and \citet{Kumar:89} found that this process gives an important contribution to the observed linewidth of  solar p-modes. It would be interesting to revisit this problem for red giants  and see what impact it has on observables like the mode visibilities, linewidths, and frequencies.

\vspace{-0.0cm}

\acknowledgements
This work was supported in part by NASA ATP grant 80NSSC21K0493.

\software{\texttt{MESA } \citep[][\url{http://mesa.sourceforge.net}]{Paxton:11, Paxton:13, Paxton:15, Paxton:18},
\texttt{GYRE } \citep[][\url{https://bitbucket.org/rhdtownsend/gyre/wiki/Home}]{Townsend:13, Townsend:18},  \texttt{SUNDIAL }\citep[\url{https://computing.llnl.gov/projects/sundials}]{Hindmarsh:05}.
}

\appendix
\vspace{-0.5cm}

\section{Coordinate transformation}
\label{sec:coord_xfm}

The step sizes in the integration of the amplitude equations are limited by the shortest timescale on which the mode amplitudes vary.   Of the four forcing terms in Equation (\ref{eq:modeampeqn}), by far the shortest timescale is set by the linear internal restoring force $i\omega_a q_a$, which induces amplitude modulations on the timescale of the mode period $2\pi/\omega_a$.  The network integrations therefore run much faster when transforming coordinates to
\beq
q_a =  q_{a,\rm lin} + x_a e^{-i\omega_a t},
\eeq
where $q_{a, \rm lin }$ is the solution to the linear equation 
\beq
\label{eq:lin_amp}
\dot{q}_{a, \rm lin} + \left(i\omega_a +\gamma_a\right)q_{a, \rm lin} = i\omega_a f_a(t).
\eeq
The equations for the nonlinear amplitudes of a parent mode $a$ coupled to a pair of daughter modes $b,c$ are then
\bea
\label{eq:nl_amp}
\dot{x}_{a} + \gamma_a x_{a} &=& 2i\omega_a \kappa_{abc} x_{b}^\ast x_{c}^\ast e^{i\Delta_{bc} t}
\non 
\dot{x}_{b} + \gamma_b x_{b} &=& 2i\omega_b \kappa_{abc} (x_{a,\rm lin}^\ast + x_{a}^\ast) x_{c}^\ast e^{i\Delta_{bc} t}
\non 
\dot{x}_{c} + \gamma_c x_{c} &=& 2i\omega_c \kappa_{abc} (x_{a,\rm lin}^\ast + x_{a}^\ast)x_{b}^\ast e^{i\Delta_{bc} t},
\eea
where $x_{a, \rm lin} = q_{a, \rm lin}e^{i\omega_a t}$, the detuning $\Delta_{bc}=\omega_a +\omega_b+\omega_c$, and we used the fact that only the parent modes are driven linearly in our calculations and thus $q_{b, \rm lin}=q_{c, \rm lin}=0$ (see Section~\ref{sec:stochastic_driving}).  Equations~(\ref{eq:nl_amp}) are easily generalized to a parent coupled to more than two daughters, a daughter coupled to more than one parent, and daughters coupled to granddaughters etc.  For a given network, we integrate Equation~(\ref{eq:lin_amp}) for each parent to determine $x_{a,\rm lin}$ and then solve Equations~(\ref{eq:nl_amp}) for the nonlinear amplitudes.  Even though Equation~(\ref{eq:lin_amp}) varies on the short timescale of the parent mode period $2\pi/\omega_a$, since there are $\lesssim50$ parent modes in each network (see Section~\ref{sec:parents}), solving for their linear amplitudes $q_{a, \rm lin}$ is relatively inexpensive.  Meanwhile, since the couplings in the networks all have small detunings $\Delta_{bc} \ll \omega_a$, the step size in Equations~(\ref{eq:nl_amp}) are set by the inverse of the damping rate or detuning rather than the much smaller mode period.
\vspace{-0.0cm}

\bibliography{refs}

\begin{thebibliography}{}
\expandafter\ifx\csname natexlab\endcsname\relax\def\natexlab#1{#1}\fi
\providecommand{\url}[1]{\href{#1}{#1}}
\providecommand{\dodoi}[1]{doi:~\href{http://doi.org/#1}{\nolinkurl{#1}}}
\providecommand{\doeprint}[1]{\href{http://ascl.net/#1}{\nolinkurl{http://ascl.net/#1}}}
\providecommand{\doarXiv}[1]{\href{https://arxiv.org/abs/#1}{\nolinkurl{https://arxiv.org/abs/#1}}}

\bibitem[{{Aerts} {et~al.}(2010){Aerts}, {Christensen-Dalsgaard}, \&
  {Kurtz}}]{Aerts:10}
{Aerts}, C., {Christensen-Dalsgaard}, J., \& {Kurtz}, D.~W. 2010,
  {Asteroseismology}

\bibitem[{{Aerts} {et~al.}(2019){Aerts}, {Mathis}, \& {Rogers}}]{Aerts:19}
{Aerts}, C., {Mathis}, S., \& {Rogers}, T.~M. 2019, \araa, 57, 35,
  \dodoi{10.1146/annurev-astro-091918-104359}

\bibitem[{{Ariaratnam} \& {Tam}(1976)}]{Ariaratnam:76}
{Ariaratnam}, S.~T., \& {Tam}, D.~S.~F. 1976, Zeitschrift Angewandte Mathematik
  und Mechanik, 56, 449, \dodoi{10.1002/zamm.19760561102}

\bibitem[{{Baglin} {et~al.}(2006){Baglin}, {Auvergne}, {Boisnard}, {Lam-Trong},
  {Barge}, {Catala}, {Deleuil}, {Michel}, \& {Weiss}}]{Baglin:06}
{Baglin}, A., {Auvergne}, M., {Boisnard}, L., {et~al.} 2006, in COSPAR Meeting,
  Vol.~36, 36th COSPAR Scientific Assembly

\bibitem[{{Ballot} {et~al.}(2011){Ballot}, {Barban}, \& {van't
  Veer-Menneret}}]{Ballot:11}
{Ballot}, J., {Barban}, C., \& {van't Veer-Menneret}, C. 2011, \aap, 531, A124,
  \dodoi{10.1051/0004-6361/201016230}

\bibitem[{{Barker} \& {Ogilvie}(2011)}]{Barker:11}
{Barker}, A.~J., \& {Ogilvie}, G.~I. 2011, \mnras, 417, 745,
  \dodoi{10.1111/j.1365-2966.2011.19322.x}

\bibitem[{{Basu} \& {Hekker}(2020)}]{Basu:20}
{Basu}, S., \& {Hekker}, S. 2020, Frontiers in Astronomy and Space Sciences, 7,
  44, \dodoi{10.3389/fspas.2020.00044}

\bibitem[{{Belkacem} {et~al.}(2011){Belkacem}, {Goupil}, {Dupret}, {Samadi},
  {Baudin}, {Noels}, \& {Mosser}}]{Belkacem:11}
{Belkacem}, K., {Goupil}, M.~J., {Dupret}, M.~A., {et~al.} 2011, \aap, 530,
  A142, \dodoi{10.1051/0004-6361/201116490}

\bibitem[{{Benomar} {et~al.}(2014){Benomar}, {Belkacem}, {Bedding}, {Stello},
  {Di Mauro}, {Ventura}, {Mosser}, {Goupil}, {Samadi}, \&
  {Garcia}}]{Benomar:14}
{Benomar}, O., {Belkacem}, K., {Bedding}, T.~R., {et~al.} 2014, \apjl, 781,
  L29, \dodoi{10.1088/2041-8205/781/2/L29}

\bibitem[{{Borucki} {et~al.}(2010){Borucki}, {Koch}, {Basri}, {Batalha},
  {Brown}, {Caldwell}, {Caldwell}, {Christensen-Dalsgaard}, {Cochran},
  {DeVore}, {Dunham}, {Dupree}, {Gautier}, {Geary}, {Gilliland}, {Gould},
  {Howell}, {Jenkins}, {Kondo}, {Latham}, {Marcy}, {Meibom}, {Kjeldsen},
  {Lissauer}, {Monet}, {Morrison}, {Sasselov}, {Tarter}, {Boss}, {Brownlee},
  {Owen}, {Buzasi}, {Charbonneau}, {Doyle}, {Fortney}, {Ford}, {Holman},
  {Seager}, {Steffen}, {Welsh}, {Rowe}, {Anderson}, {Buchhave}, {Ciardi},
  {Walkowicz}, {Sherry}, {Horch}, {Isaacson}, {Everett}, {Fischer}, {Torres},
  {Johnson}, {Endl}, {MacQueen}, {Bryson}, {Dotson}, {Haas}, {Kolodziejczak},
  {Van Cleve}, {Chandrasekaran}, {Twicken}, {Quintana}, {Clarke}, {Allen},
  {Li}, {Wu}, {Tenenbaum}, {Verner}, {Bruhweiler}, {Barnes}, \&
  {Prsa}}]{Borucki:10}
{Borucki}, W.~J., {Koch}, D., {Basri}, G., {et~al.} 2010, Science, 327, 977,
  \dodoi{10.1126/science.1185402}

\bibitem[{{Chang} \& {Gough}(1998)}]{Chang:98}
{Chang}, H.-Y., \& {Gough}, D.~O. 1998, \solphys, 181, 251,
  \dodoi{10.1023/A:1005017817714}

\bibitem[{{Christensen-Dalsgaard}(2002)}]{CD:02}
{Christensen-Dalsgaard}, J. 2002, Reviews of Modern Physics, 74, 1073,
  \dodoi{10.1103/RevModPhys.74.1073}

\bibitem[{{Christensen-Dalsgaard} {et~al.}(1989){Christensen-Dalsgaard},
  {Gough}, \& {Libbrecht}}]{CD:89}
{Christensen-Dalsgaard}, J., {Gough}, D.~O., \& {Libbrecht}, K.~G. 1989, \apjl,
  341, L103, \dodoi{10.1086/185468}

\bibitem[{{Corsaro} {et~al.}(2015){Corsaro}, {De Ridder}, \&
  {Garc{\'{\i}}a}}]{Corsaro:15}
{Corsaro}, E., {De Ridder}, J., \& {Garc{\'{\i}}a}, R.~A. 2015, \aap, 579, A83,
  \dodoi{10.1051/0004-6361/201525895}

\bibitem[{{Deheuvels} {et~al.}(2012){Deheuvels}, {Garc{\'{\i}}a}, {Chaplin},
  {Basu}, {Antia}, {Appourchaux}, {Benomar}, {Davies}, {Elsworth}, {Gizon},
  {Goupil}, {Reese}, {Regulo}, {Schou}, {Stahn}, {Casagrande},
  {Christensen-Dalsgaard}, {Fischer}, {Hekker}, {Kjeldsen}, {Mathur}, {Mosser},
  {Pinsonneault}, {Valenti}, {Christiansen}, {Kinemuchi}, \&
  {Mullally}}]{Deheuvels:12}
{Deheuvels}, S., {Garc{\'{\i}}a}, R.~A., {Chaplin}, W.~J., {et~al.} 2012, \apj,
  756, 19, \dodoi{10.1088/0004-637X/756/1/19}

\bibitem[{{Deheuvels} {et~al.}(2014){Deheuvels}, {Do{\u g}an}, {Goupil},
  {Appourchaux}, {Benomar}, {Bruntt}, {Campante}, {Casagrande}, {Ceillier},
  {Davies}, {De Cat}, {Fu}, {Garc{\'{\i}}a}, {Lobel}, {Mosser}, {Reese},
  {Regulo}, {Schou}, {Stahn}, {Thygesen}, {Yang}, {Chaplin},
  {Christensen-Dalsgaard}, {Eggenberger}, {Gizon}, {Mathis},
  {Molenda-{\.Z}akowicz}, \& {Pinsonneault}}]{Deheuvels:14}
{Deheuvels}, S., {Do{\u g}an}, G., {Goupil}, M.~J., {et~al.} 2014, \aap, 564,
  A27, \dodoi{10.1051/0004-6361/201322779}

\bibitem[{{Dupret} {et~al.}(2009){Dupret}, {Belkacem}, {Samadi}, {Montalban},
  {Moreira}, {Miglio}, {Godart}, {Ventura}, {Ludwig}, {Grigahc{\`e}ne},
  {Goupil}, {Noels}, \& {Caffau}}]{Dupret:09}
{Dupret}, M.-A., {Belkacem}, K., {Samadi}, R., {et~al.} 2009, \aap, 506, 57,
  \dodoi{10.1051/0004-6361/200911713}

\bibitem[{{Essick} \& {Weinberg}(2016)}]{Essick:16}
{Essick}, R., \& {Weinberg}, N.~N. 2016, \apj, 816, 18,
  \dodoi{10.3847/0004-637X/816/1/18}

\bibitem[{{Fuller} {et~al.}(2015){Fuller}, {Cantiello}, {Stello}, {Garcia}, \&
  {Bildsten}}]{Fuller:15}
{Fuller}, J., {Cantiello}, M., {Stello}, D., {Garcia}, R.~A., \& {Bildsten}, L.
  2015, Science, 350, 423, \dodoi{10.1126/science.aac6933}

\bibitem[{{Goldreich} \& {Kumar}(1988)}]{Goldreich:88}
{Goldreich}, P., \& {Kumar}, P. 1988, \apj, 326, 462, \dodoi{10.1086/166108}

\bibitem[{{Grosjean} {et~al.}(2014){Grosjean}, {Dupret}, {Belkacem},
  {Montalban}, {Samadi}, \& {Mosser}}]{Grosjean:14}
{Grosjean}, M., {Dupret}, M.-A., {Belkacem}, K., {et~al.} 2014, \aap, 572, A11,
  \dodoi{10.1051/0004-6361/201423827}

\bibitem[{{Hekker} \& {Christensen-Dalsgaard}(2017)}]{Hekker:17}
{Hekker}, S., \& {Christensen-Dalsgaard}, J. 2017, \aapr, 25, 1,
  \dodoi{10.1007/s00159-017-0101-x}

\bibitem[{Hindmarsh {et~al.}(2005)Hindmarsh, Brown, Grant, Lee, Serban,
  Shumaker, \& Woodward}]{Hindmarsh:05}
Hindmarsh, A.~C., Brown, P.~N., Grant, K.~E., {et~al.} 2005, ACM Transactions
  on Mathematical Software (TOMS), 31, 363

\bibitem[{{Huber} {et~al.}(2010){Huber}, {Bedding}, {Stello}, {Mosser},
  {Mathur}, {Kallinger}, {Hekker}, {Elsworth}, {Buzasi}, {De Ridder},
  {Gilliland}, {Kjeldsen}, {Chaplin}, {Garc{\'{\i}}a}, {Hale}, {Preston},
  {White}, {Borucki}, {Christensen-Dalsgaard}, {Clarke}, {Jenkins}, \&
  {Koch}}]{Huber:10}
{Huber}, D., {Bedding}, T.~R., {Stello}, D., {et~al.} 2010, \apj, 723, 1607,
  \dodoi{10.1088/0004-637X/723/2/1607}

\bibitem[{{Kjeldsen} \& {Bedding}(1995)}]{Kjeldsen:95}
{Kjeldsen}, H., \& {Bedding}, T.~R. 1995, \aap, 293, 87

\bibitem[{{Kumar} {et~al.}(1988){Kumar}, {Franklin}, \& {Goldreich}}]{Kumar:88}
{Kumar}, P., {Franklin}, J., \& {Goldreich}, P. 1988, \apj, 328, 879,
  \dodoi{10.1086/166345}

\bibitem[{{Kumar} \& {Goldreich}(1989)}]{Kumar:89}
{Kumar}, P., \& {Goldreich}, P. 1989, \apj, 342, 558, \dodoi{10.1086/167616}

\bibitem[{{Kumar} {et~al.}(1994){Kumar}, {Goldreich}, \& {Kerswell}}]{Kumar:94}
{Kumar}, P., {Goldreich}, P., \& {Kerswell}, R. 1994, \apj, 427, 483,
  \dodoi{10.1086/174159}

\bibitem[{{Kumar} \& {Goodman}(1996)}]{Kumar:96}
{Kumar}, P., \& {Goodman}, J. 1996, \apj, 466, 946, \dodoi{10.1086/177565}

\bibitem[{{Loi} \& {Papaloizou}(2018)}]{Loi:18}
{Loi}, S.~T., \& {Papaloizou}, J. C.~B. 2018, \mnras, 477, 5338,
  \dodoi{10.1093/mnras/sty917}

\bibitem[{{Mosser} {et~al.}(2011){Mosser}, {Barban}, {Montalb{\'a}n}, {Beck},
  {Miglio}, {Belkacem}, {Goupil}, {Hekker}, {De Ridder}, {Dupret}, {Elsworth},
  {Noels}, {Baudin}, {Michel}, {Samadi}, {Auvergne}, {Baglin}, \&
  {Catala}}]{Mosser:11}
{Mosser}, B., {Barban}, C., {Montalb{\'a}n}, J., {et~al.} 2011, \aap, 532, A86,
  \dodoi{10.1051/0004-6361/201116825}

\bibitem[{{Mosser} {et~al.}(2012{\natexlab{a}}){Mosser}, {Elsworth}, {Hekker},
  {Huber}, {Kallinger}, {Mathur}, {Belkacem}, {Goupil}, {Samadi}, {Barban},
  {Bedding}, {Chaplin}, {Garc{\'{\i}}a}, {Stello}, {De Ridder}, {Middour},
  {Morris}, \& {Quintana}}]{Mosser:12:depressed}
{Mosser}, B., {Elsworth}, Y., {Hekker}, S., {et~al.} 2012{\natexlab{a}}, \aap,
  537, A30, \dodoi{10.1051/0004-6361/201117352}

\bibitem[{{Mosser} {et~al.}(2012{\natexlab{b}}){Mosser}, {Goupil}, {Belkacem},
  {Marques}, {Beck}, {Bloemen}, {De Ridder}, {Barban}, {Deheuvels}, {Elsworth},
  {Hekker}, {Kallinger}, {Ouazzani}, {Pinsonneault}, {Samadi}, {Stello},
  {Garc{\'{\i}}a}, {Klaus}, {Li}, {Mathur}, \& {Morris}}]{Mosser:12:rotation}
{Mosser}, B., {Goupil}, M.~J., {Belkacem}, K., {et~al.} 2012{\natexlab{b}},
  \aap, 548, A10, \dodoi{10.1051/0004-6361/201220106}

\bibitem[{{Mosser} {et~al.}(2017){Mosser}, {Belkacem}, {Pin{\c c}on}, {Takata},
  {Vrard}, {Barban}, {Goupil}, {Kallinger}, \& {Samadi}}]{Mosser:17}
{Mosser}, B., {Belkacem}, K., {Pin{\c c}on}, C., {et~al.} 2017, \aap, 598, A62,
  \dodoi{10.1051/0004-6361/201629494}

\bibitem[{{Paxton} {et~al.}(2011){Paxton}, {Bildsten}, {Dotter}, {Herwig},
  {Lesaffre}, \& {Timmes}}]{Paxton:11}
{Paxton}, B., {Bildsten}, L., {Dotter}, A., {et~al.} 2011, \apjs, 192, 3,
  \dodoi{10.1088/0067-0049/192/1/3}

\bibitem[{{Paxton} {et~al.}(2013){Paxton}, {Cantiello}, {Arras}, {Bildsten},
  {Brown}, {Dotter}, {Mankovich}, {Montgomery}, {Stello}, {Timmes}, \&
  {Townsend}}]{Paxton:13}
{Paxton}, B., {Cantiello}, M., {Arras}, P., {et~al.} 2013, \apjs, 208, 4,
  \dodoi{10.1088/0067-0049/208/1/4}

\bibitem[{{Paxton} {et~al.}(2015){Paxton}, {Marchant}, {Schwab}, {Bauer},
  {Bildsten}, {Cantiello}, {Dessart}, {Farmer}, {Hu}, {Langer}, {Townsend},
  {Townsley}, \& {Timmes}}]{Paxton:15}
{Paxton}, B., {Marchant}, P., {Schwab}, J., {et~al.} 2015, \apjs, 220, 15,
  \dodoi{10.1088/0067-0049/220/1/15}

\bibitem[{{Paxton} {et~al.}(2018){Paxton}, {Schwab}, {Bauer}, {Bildsten},
  {Blinnikov}, {Duffell}, {Farmer}, {Goldberg}, {Marchant}, {Sorokina},
  {Thoul}, {Townsend}, \& {Timmes}}]{Paxton:18}
{Paxton}, B., {Schwab}, J., {Bauer}, E.~B., {et~al.} 2018, \apjs, 234, 34,
  \dodoi{10.3847/1538-4365/aaa5a8}

\bibitem[{{Poulin} \& {Flierl}(2008)}]{Poulin:08}
{Poulin}, F.~J., \& {Flierl}, G.~R. 2008, Proceedings of the Royal Society of
  London Series A, 464, 1885, \dodoi{10.1098/rspa.2008.0007}

\bibitem[{{Press}(1981)}]{Press:81}
{Press}, W.~H. 1981, \apj, 245, 286, \dodoi{10.1086/158809}

\bibitem[{{Ricker} {et~al.}(2015){Ricker}, {Winn}, {Vanderspek}, {Latham},
  {Bakos}, {Bean}, {Berta-Thompson}, {Brown}, {Buchhave}, \&
  {Butler}}]{Ricker:15}
{Ricker}, G.~R., {Winn}, J.~N., {Vanderspek}, R., {et~al.} 2015, Journal of
  Astronomical Telescopes, Instruments, and Systems, 1, 014003,
  \dodoi{10.1117/1.JATIS.1.1.014003}

\bibitem[{{Samadi}(2011)}]{Samadi:11}
{Samadi}, R. 2011, in Lecture Notes in Physics, Berlin Springer Verlag, Vol.
  832, Lecture Notes in Physics, Berlin Springer Verlag, ed. J.-P. {Rozelot} \&
  C.~{Neiner}, 305, \dodoi{10.1007/978-3-642-19928-8_11}

\bibitem[{{Samadi} {et~al.}(2012){Samadi}, {Belkacem}, {Dupret}, {Ludwig},
  {Baudin}, {Caffau}, {Goupil}, \& {Barban}}]{Samadi:12}
{Samadi}, R., {Belkacem}, K., {Dupret}, M.~A., {et~al.} 2012, \aap, 543, A120,
  \dodoi{10.1051/0004-6361/201219253}

\bibitem[{{Samadi} {et~al.}(2007){Samadi}, {Georgobiani}, {Trampedach},
  {Goupil}, {Stein}, \& {Nordlund}}]{Samadi:07}
{Samadi}, R., {Georgobiani}, D., {Trampedach}, R., {et~al.} 2007, \aap, 463,
  297, \dodoi{10.1051/0004-6361:20041953}

\bibitem[{{Schenk} {et~al.}(2002){Schenk}, {Arras}, {Flanagan}, {Teukolsky}, \&
  {Wasserman}}]{Schenk:02}
{Schenk}, A.~K., {Arras}, P., {Flanagan}, {\'E}.~{\'E}., {Teukolsky}, S.~A., \&
  {Wasserman}, I. 2002, \prd, 65, 024001, \dodoi{10.1103/PhysRevD.65.024001}

\bibitem[{{Stello} {et~al.}(2016{\natexlab{a}}){Stello}, {Cantiello}, {Fuller},
  {Garcia}, \& {Huber}}]{Stello:16a}
{Stello}, D., {Cantiello}, M., {Fuller}, J., {Garcia}, R.~A., \& {Huber}, D.
  2016{\natexlab{a}}, \pasa, 33, e011, \dodoi{10.1017/pasa.2016.9}

\bibitem[{{Stello} {et~al.}(2016{\natexlab{b}}){Stello}, {Cantiello}, {Fuller},
  {Huber}, {Garc{\'{\i}}a}, {Bedding}, {Bildsten}, \& {Silva
  Aguirre}}]{Stello:16b}
{Stello}, D., {Cantiello}, M., {Fuller}, J., {et~al.} 2016{\natexlab{b}}, \nat,
  529, 364, \dodoi{10.1038/nature16171}

\bibitem[{{Stello} {et~al.}(2009){Stello}, {Chaplin}, {Basu}, {Elsworth}, \&
  {Bedding}}]{Stello:09}
{Stello}, D., {Chaplin}, W.~J., {Basu}, S., {Elsworth}, Y., \& {Bedding}, T.~R.
  2009, \mnras, 400, L80, \dodoi{10.1111/j.1745-3933.2009.00767.x}

\bibitem[{Stratonovich \& Romanovskii(1965)}]{Stratonovich:65}
Stratonovich, R.~L., \& Romanovskii, Y.~M. 1965, in Non-Linear Transformations
  of Stochastic Processes, ed. P.~Kuznetsov, R.~Stratonovich, \& V.~Tikhonov
  (Pergamon), 327 -- 338

\bibitem[{Townsend {et~al.}(2018)Townsend, Goldstein, \& Zweibel}]{Townsend:18}
Townsend, R. H.~D., Goldstein, J., \& Zweibel, E.~G. 2018, Monthly Notices of
  the Royal Astronomical Society, 475, 879, \dodoi{10.1093/mnras/stx3142}

\bibitem[{{Townsend} \& {Teitler}(2013)}]{Townsend:13}
{Townsend}, R.~H.~D., \& {Teitler}, S.~A. 2013, \mnras, 435, 3406,
  \dodoi{10.1093/mnras/stt1533}

\bibitem[{{Van Hoolst}(1994)}]{VanHoolst:94}
{Van Hoolst}, T. 1994, aap, 286, 879

\bibitem[{{van Kampen}(1992)}]{vanKampen:92}
{van Kampen}, N.~G. 1992, Stochastic Processes in Physics and Chemistry
  Publisher: Elsevier Science, Amsterdam, 1992

\bibitem[{{Vrard} {et~al.}(2018){Vrard}, {Kallinger}, {Mosser}, {Barban},
  {Baudin}, {Belkacem}, \& {Cunha}}]{Vrard:18}
{Vrard}, M., {Kallinger}, T., {Mosser}, B., {et~al.} 2018, \aap, 616, A94,
  \dodoi{10.1051/0004-6361/201732477}

\bibitem[{{Weinberg} \& {Arras}(2019)}]{Weinberg:19}
{Weinberg}, N.~N., \& {Arras}, P. 2019, \apj, 873, 67,
  \dodoi{10.3847/1538-4357/ab0204}

\bibitem[{{Weinberg} {et~al.}(2012){Weinberg}, {Arras}, {Quataert}, \&
  {Burkart}}]{Weinberg:12}
{Weinberg}, N.~N., {Arras}, P., {Quataert}, E., \& {Burkart}, J. 2012, \apj,
  751, 136, \dodoi{10.1088/0004-637X/751/2/136}

\bibitem[{{Wu} \& {Goldreich}(2001)}]{Wu:01}
{Wu}, Y., \& {Goldreich}, P. 2001, \apj, 546, 469, \dodoi{10.1086/318234}

\bibitem[{{Yu} {et~al.}(2020){Yu}, {Weinberg}, \& {Fuller}}]{Yu:20}
{Yu}, H., {Weinberg}, N.~N., \& {Fuller}, J. 2020, \mnras, 496, 5482,
  \dodoi{10.1093/mnras/staa1858}

\bibitem[{{Zhang} {et~al.}(1993){Zhang}, {Casademunt}, \&
  {Vi{\~n}als}}]{Zhang:93}
{Zhang}, W., {Casademunt}, J., \& {Vi{\~n}als}, J. 1993, Physics of Fluids A,
  5, 3147, \dodoi{10.1063/1.858723}

\end{thebibliography}
\end{document}